\documentclass[pra,twocolumn,superscriptaddress,floatfix,nofootinbib]{revtex4-2}
\pdfoutput=1 

\usepackage[T1]{fontenc}
\usepackage{amsmath}
\usepackage{amssymb}
\usepackage{amsfonts}
\usepackage{upgreek}
\usepackage{graphicx}
\usepackage{enumitem}
\usepackage{bm}
\usepackage{rotating}
\usepackage{hyperref}
\usepackage{pbox}
\usepackage[usenames,dvipsnames]{xcolor}
\usepackage[percent]{overpic}
\usepackage{physics}
\usepackage{dsfont}
\usepackage{mathtools}
\usepackage{leftidx}
\usepackage{lmodern}
\usepackage[normalem]{ulem}
\usepackage{braket}
\usepackage{tensor}
\usepackage{mathrsfs}
\usepackage{float}
\usepackage{dsfont}
\usepackage[margin=2cm]{geometry}
\usepackage[title]{appendix}
\usepackage{tcolorbox}
\usepackage{accents}

\makeatletter
\renewcommand{\l@subsection}[2]{}
\makeatother

\newcommand{\exd}{{\hbox{d}}}

\def\trB{\underset{\text{E}}{\text{Tr}}}

\begin{document}

\title{How I Learned to Stop Worrying and Love the Redfield Equation: Completely-Positive Resummation of Non-Markovian Dynamics}

\author{Greg Kaplanek}
\email{gkaplane@syr.edu}
\affiliation{Institute for Quantum \& Information Sciences, Syracuse University, NY 13210, USA}
\affiliation{Department of Electrical Engineering and Computer Science, Syracuse University, NY 13210, USA}

\author{Jason Pollack}
\email{japollac@syr.edu}
\affiliation{Institute for Quantum \& Information Sciences, Syracuse University, NY 13210, USA}
\affiliation{Department of Electrical Engineering and Computer Science, Syracuse University, NY 13210, USA}

\begin{abstract} 

The Redfield equation with time-dependent coefficients (or TCL$_2$ equation) gives one of the simplest perturbative descriptions of open quantum system dynamics derived from a microscopic theory. Its status as an approximation is nevertheless subtle: while perturbation theory fixes the reduced dynamics order-by-order in the system-environment coupling, solving the Redfield equation yields a particular resummation that need not preserve complete positivity. Here we compare Redfield evolution with an exponentiated (Magnus) resummation of the same second-order perturbative map, which is completely positive and trace preserving (CPTP) for a broad class of physical conditions. We also construct a simple CPTP resolvent resummation, illustrating the non-uniqueness of resummations of the same perturbative map. The Redfield and Magnus prescriptions agree to second order but differ through time-ordering corrections at higher orders. We exhibit a weak-coupling criterion which guarantees that Redfield evolution, even when non-Markovian, remains close to a CPTP map. We illustrate the criterion in two examples, focusing on finite-dimensional (qubit) systems. Our results clarify when, and why, Redfield evolution can remain reliable even when it is highly non-Lindbladian.

\end{abstract}

\maketitle

\tableofcontents

\section{Introduction}

The Redfield equation is one of the oldest and simplest perturbative descriptions of open quantum system dynamics derived from a microscopic theory \cite{redfield1957theory,redfield1965theory}. It is also one of the most mistrusted, for two main reasons. First, although the Redfield equation follows directly from weak-coupling perturbation theory, its instantaneous generator is generically not of Gorini-Kossakowski-Lindblad-Sudarshan (GKLS) form \cite{Lindblad:1975ef,Gorini:1976cm}. Second, Redfield evolution has been shown in various examples to exhibit violations of positivity or complete positivity. This has led to a long-standing view that Redfield evolution should either be used with considerable caution or supplemented by further approximations \cite{suarez1992memory,gaspard1999slippage,Benatti:2005uq,zhao2002density,Jeske:2015mlq,Benatti:2022ect,delaPradilla:2024cnp,Tokieda:2023yig,Tupkary:2023txe}.

The purpose of this paper is to revisit this conclusion from a different perspective. We will take the point of view that the issue is not perturbation theory itself, but what one chooses to do with it. Suppose that the exact reduced dynamics of an open system are described by the quantum map
\begin{equation}
\varrho_{\mathrm{S}}(t) = \Phi(t,t_{\mathrm{i}})[\varrho_{\mathrm{S}}(t_{\mathrm{i}}) ],
\end{equation}
where $\Phi(t,t_{\mathrm{i}})$ is a `propagator:' an exact, completely-positive and trace-preserving (CPTP) quantum channel implementing the time evolution from $t_i$ to $t$. Suppose further that microscopic perturbation theory determines this map through second order in the system-environment coupling $\lambda$,
\begin{equation}  \Phi(t,t_{\mathrm i}) \simeq \mathcal I
+
\lambda^2\Phi_2(t,t_{\mathrm{i}})
+
\ldots
\label{IntroPertMap}
\end{equation}
As we discuss below, a fixed-order approximation such as Eq.~(\ref{IntroPertMap}) need not itself remain CPTP, even though the exact map $\Phi(t,t_{\mathrm i})$ is CPTP by construction. 

Eq.~(\ref{IntroPertMap}) fixes the dynamics only up to  second order, which means that any prescription or {\it resummation} promoting it to a finite expression valid beyond this order necessarily chooses an infinite collection of higher-order terms. In this sense, resummation is an art: different resummations agree with the same perturbative calculation through $\mathcal O(\lambda^2)$, but can have very different behaviour once the higher-order terms become important. One may choose these terms to preserve complete positivity, or instead choose to capture some other feature of the underlying late-time dynamics. Perturbation theory by itself up to a fixed order does not uniquely select among these choices.

The Redfield equation with time-dependent coefficients (Redfield TDC), equivalently the second-order time-convolutionless equation (TCL$_2$) \cite{shibata1977generalized,chaturvedi1979time,shibata1980expansion}, fixes one such resummation. For the microscopic interaction
\begin{equation} \label{Hint_introduction}
\lambda H_{\mathrm{int}}^I(t)
=
\lambda \sum_\alpha S_\alpha(t)\otimes E_\alpha(t) \; ,
\end{equation}
it takes the form
\begin{eqnarray} \label{Red_H_centered_introduction}
\frac{\exd \varrho_{\mathrm{S}}}{\exd t} & \simeq & \lambda^2 \sum_{\alpha,\beta}
\int_{t_{\mathrm i}}^{t}\exd s \, 
\bigg( \,  [S_\alpha(s)\varrho_{\mathrm{S}}(t),S_\beta(t)]
\, \mathcal E_{\beta\alpha}(t,s) \\
&& \hspace{19.8mm} 
+ [S_\beta(t),\varrho_{\mathrm{S}}(t) S_\alpha(s)] \, 
\mathcal E_{\beta\alpha}^{*}(t,s) \, \bigg)\ , \notag 
\end{eqnarray}
with $S_\alpha(t)$ and $E_\alpha(t)$ Hermitian operators on the system and environment, respectively. We assume a factorized initial state $\varrho_{\mathrm{Si}} \otimes \varrho_{\mathrm{Ei}}$ at $t=t_{\mathrm{i}}$ and, for simplicity, vanishing environment
one-point functions\footnote{This assumption is made in the main body in Eq.~(\ref{MFvanish})
for simplicity, but the arguments of this paper extend straightforwardly to non-vanishing one-point functions; see Appendix~\ref{App:MF}.},
\begin{equation}
\trB
\big[
E_\alpha(t)
\rho_{\mathrm{Ei}}
\big] = 0 \; , \label{MFvanish_introduction}
\end{equation}
with environment correlation functions
\begin{equation}
\mathcal E_{\alpha\beta}(t,s)
=
\trB
\big[
E_\alpha(t)
E_\beta(s)
\rho_{\mathrm{Ei}}
\big] \; . \label{plaindefEenv_introduction}
\end{equation}
We review the derivation of Eq.~(\ref{Red_H_centered_introduction}) and some of its known subtleties in Sec.~\ref{sec:Redfield}. Throughout this work we retain the finite limits on the integrals appearing on the right-hand side of Eq.~(\ref{Red_H_centered_introduction}), and refer to this as the Redfield equation with time-dependent coefficients. This distinguishes it from the asymptotic Redfield equation, obtained by taking $t-t_{\mathrm i}\to\infty$ in the corresponding integrals.

In the notation of Eq.~(\ref{IntroPertMap}), the Redfield TDC takes the particularly simple form 
\begin{equation}
\frac{\exd\varrho_{\mathrm S}}{\exd t} = \lambda^2\dot{\Phi}_2(t,t_{\mathrm{i}})
\big[
\varrho_{\mathrm S}(t)
\big].
\label{IntroRedfield}
\end{equation}
This direct promotion of the perturbative generator to a local master equation is precisely what makes Redfield evolution such a natural resummation of the microscopic calculation. It has the formal solution $\varrho_{\mathrm{S}}(t) = \Phi_{\mathrm{Red}}(t,t_{\mathrm{i}})[\varrho_{\mathrm{S}}(t_{\mathrm{i}}) ]$, where
\begin{equation}
\Phi_{\rm Red}(t,t_{\mathrm i})
 = \mathcal T e^{
\lambda^2
\int_{t_{\mathrm i}}^t
\mathrm{d} s \; 
\dot{\Phi}_2(s,t_{\mathrm{i}}) } .\label{IntroRedfieldMap}
\end{equation}
Here $\mathcal T$ denotes the standard time-ordering prescription on superoperators (see Appendix \ref{App:Magnus}).

The resummation $\Phi_{\rm Red}(t,t_{\mathrm i})$ agrees with the result of ordinary perturbation theory in Eq.~\eqref{IntroPertMap} up through terms of order $\mathcal O(\lambda^2)$, but the time-ordered exponential generates contributions at every higher order in $\lambda$. Explicitly expanding the above, we find
\begin{eqnarray}
&& \Phi_{\rm Red}(t,t_{\mathrm i}) \; \simeq \; \mathcal{I} + \lambda^2 \Phi_{2}(t,t_{\mathrm{i}}) \label{Intro_PhiRed_Exp}\\
& & \qquad + \lambda^4 \int_{t_{\mathrm{i}}}^{t} \exd s_1 \int_{t_{\mathrm{i}}}^{s_{1}} \exd s_2  \; \dot{\Phi}_{2}(s_{1},t_{\mathrm{i}})  \dot{\Phi}_{2}(s_{2},t_{\mathrm{i}}) + \ldots \notag 
\end{eqnarray}
The higher-order contributions here are not fixed by the second-order microscopic calculation: they are the particular resummation specified by the Redfield TDC. 
Although $\dot{\Phi}_2$ is generically not of GKLS form, this alone does not determine whether the resulting finite-time Redfield map $\Phi_{\rm Red}$ is CPTP.

A second-order master equation such as Eq.~(\ref{IntroRedfield}) does not, in general, determine the reduced state to higher orders in $\lambda$: obtaining the state consistently to a given perturbative order may require the generator at still higher order \cite{Fleming:2011irq}. Higher-order TCL expansions have therefore received renewed attention \cite{Chen:2025mgm,Suarez:2025bhb,Kumar:2025vel,Lampert:2025cus}. Our aim here is complementary. Rather than computing the exact higher-order generator, we ask what can already be learned by comparing different resummations of the same second-order microscopic result $\Phi_2(t,t_{\mathrm{i}})$. In particular, we compare Redfield evolution with alternative resummations of $\Phi_2(t,t_{\mathrm{i}})$ that preserve complete positivity by construction.

In our analysis we assume a positivity property of the environmental correlator, which implies that it admits a mode decomposition of the form
\begin{equation}
\mathcal E_{\alpha\beta}(t,s)
=
\sum_\kappa
u_{\kappa\alpha}(t)
u_{\kappa\beta}^{\ast}(s)\ .
\label{envCORR_intro}
\end{equation}
We discuss this property in detail in \S\ref{sec:PT}, but emphasize here that it arises naturally from the microscopic definition of $\mathcal{E}(t,s)$ as a quantum correlation function.

Crucially, with the extra assumption of the decomposition in Eq.~(\ref{envCORR_intro}), the second-order perturbative correction $\Phi_2(t)$ itself now has a GKLS-like structure when the physical time $t$ is held fixed. This immediately gives many possible CPTP resummations. Two examples we consider below are a resolvent resummation,
\begin{equation} 
\Phi_{\rm res}(t,t_{\mathrm{i}}) = 
\big(
\mathcal I-\lambda^2\Phi_2(t,t_{\mathrm{i}})
\big)^{-1}.
\label{IntroCPTPresum}
\end{equation}
and a Magnus resummation,
\begin{equation}
\Phi_{\rm Mag}(t,t_{\mathrm{i}}) = 
e^{\lambda^2\Phi_2(t,t_{\mathrm{i}})}.
\end{equation}
Both choices of resummation reproduce Eq.~(\ref{IntroPertMap}) through $\mathcal O(\lambda^2)$, and both are CPTP under the positivity condition above, but they make different choices at $\mathcal O(\lambda^4)$ and beyond. 

In particular, the Magnus map has the expansion ({\it cf.}~Eq.(\ref{Intro_PhiRed_Exp}))
\begin{equation}
\Phi_{\rm Mag}(t,t_{\mathrm i}) \;  \simeq  \; \mathcal{I} + \lambda^2 \Phi_{2}(t,t_{\mathrm{i}}) + \frac{\lambda^4}{2} \big( \Phi_{2}(t,t_{\mathrm{i}}) \big)^2 + \ldots   \label{Intro_PhiMag_Exp}
\end{equation}
The Magnus map has already appeared in the literature: the same exponentiated second-order map appears in the cumulant equation of Alicki \cite{alicki1989master}, dynamical coarse-graining of Schaller and Brandes \cite{Schaller:2008jiu}, and the refined weak-coupling expansion of Rivas \cite{rivas2017refined,Rivas:2019wug}. Here we arrive at the map as the leading term of the Magnus expansion of the time-ordered Redfield evolution. Our construction also requires only positivity of $\mathcal{E}_{\alpha\beta}$, not the time-translation invariance assumed in these earlier works.

None of the maps $\Phi_{\mathrm{res}}$, $\Phi_{\mathrm{Mag}}$, or $\Phi_{\mathrm{Red}}$ is uniquely ``the'' correct resummation. There is nevertheless a practical reason to single out the Redfield map: it follows directly by truncating the microscopic time-local generator at second order and solving the resulting master equation. It is therefore natural to ask when this simple perturbatively-derived evolution remains close to a manifestly CPTP completion of the same second-order data. The Magnus map is particularly useful for answering this question, since its difference from Redfield is entirely controlled by the superoperator time ordering in Eq.~(\ref{Intro_PhiRed_Exp}). We derive a simple sufficient criterion for $\Phi_{\mathrm{Red}}\simeq\Phi_{\mathrm{Mag}}$ and test it in several examples. This criterion does not by itself guarantee agreement with the exact dynamics. Rather, it tells us when the Redfield evolution --- the natural master equation obtained by truncating the perturbatively derived time-local generator --- remains close to a manifestly CPTP resummation of the same second-order data.

\vspace{3mm}

The remainder of the manuscript is organized as follows. \S~\ref{Sec:Lindblad} reviews the distinction between GKLS evolution, CP divisibility, and finite-time complete positivity. In \S~\ref{sec:PT}, we show that the second-order perturbative map admits an approximate Kraus representation whenever Eq.~(\ref{envCORR_intro}) holds, and discuss how complete positivity can nevertheless fail at fixed perturbative order. \S~\ref{Sec:Resolvent} uses (\ref{envCORR_intro}) to rewrite the second-order map in a GKLS-like form uses this to construct the simple CPTP resolvent resummation $\Phi_{\mathrm{res}}$, while \S~\ref{sec:Redfield} interprets the Redfield TDC as a different, time-local exponentiated resummation of the same second-order map. In \S~\ref{sec:Magnus}, we compare Redfield with the CPTP Magnus resummation $\Phi_{\mathrm{Mag}}$ and derive a criterion for when superoperator time-ordering corrections are small.

We apply these results to a qubit coupled to an arbitrary environment in \S~\ref{sec:qubitCP}, before studying two explicit environments: a single qubit in \S~\ref{sec:qubitqubit} and a bath of harmonic oscillators in \S~\ref{sec:qubit_damp}. Appendix~\ref{App:Magnus} reviews the relation between time-ordered and Magnus expansions, while Appendix~\ref{App:MF} extends our results to non-vanishing environmental one-point functions.

\section{The usual story: GKLS evolution}
\label{Sec:Lindblad}

GKLS evolution provides the canonical framework in which complete positivity is manifest \cite{Lindblad:1975ef,Gorini:1976cm}. In its original time-independent form, it generates a CPTP quantum dynamical semigroup, while time-dependent generalizations lead naturally to the notion of CP-divisible dynamics \cite{Alicki:1979wxy,davies1978open,Breuer:2004cdm,Breuer:2009pku,Rivas:2010pzm,Plenio:2014nyj}. It is necessary to distinguish these properties carefully, since the finite-time complete positivity of a reduced map requires neither semigroup structure nor CP divisibility.

Consider first a time-independent master equation
\begin{equation}
\frac{\partial \varrho_{\mathrm{S}}}{\partial t}
= \mathcal{L}\big( \varrho_{\mathrm{S}}(t) \big),
\label{GKLS_eq}
\end{equation}
with
\begin{eqnarray}
\mathcal{L}\big( \varrho_{\mathrm{S}}(t) \big) & = & 
- i
\big[
H_{\mathrm{S}} + \Delta H,
\varrho_{\mathrm{S}}
\big] \label{GKLS_generator} \\
&& +
\sum_{\alpha,\beta}
h_{\alpha \beta}
\bigg(
O_{\alpha}\,
\varrho_{\mathrm{S}}\,
O_{\beta}^{\dagger}
-
\frac{1}{2}
\Big\{
O_{\beta}^{\dagger} \, O_{\alpha},
\varrho_{\mathrm{S}}
\Big\}
\bigg) \; . \notag
\end{eqnarray}
Here $O_\alpha$ are jump operators, $H_{\mathrm S}$ is the system
Hamiltonian, $\Delta H$ is the environment-induced effective Hamiltonian (sometimes called a Lamb shift), and $h=[h_{\alpha\beta}]$ is the Kossakowski matrix. If the matrix $h$ is positive semi-definite, the generator is of GKLS (or Lindblad) form; equivalently, one may diagonalize $h$ and express the dissipative part in terms of non-negative, time-independent rates.

The results of Lindblad \cite{Lindblad:1975ef} and Gorini, Kossakowski \& Sudarshan \cite{Gorini:1976cm} establish that such a generator produces the CPTP map
\begin{equation}
\varrho_{\mathrm{S}}(t)
=
\Phi(t,t_{\mathrm{i}})
\big[
\varrho_{\mathrm{S}}(t_{\mathrm{i}})
\big] \quad \mathrm{with}
\ \ 
\Phi(t,t_{\mathrm{i}})
=
e^{(t-t_{\mathrm{i}})\mathcal{L}} ,
\end{equation}
where $\mathcal{L}$ is the time-independent generator from Eq.~(\ref{GKLS_generator}). Moreover, the evolution forms a quantum dynamical semigroup such that
\begin{equation} \label{semigroup}
\Phi(t_3,t_1) = \Phi(t_3,t_2) \,  
\Phi(t_2,t_1)
\end{equation}
which is often taken to be synonymous with Markovianity.

There is a rigorous microscopic sense in which precisely this structure
emerges at weak coupling. The classic results of Davies \cite{davies1974markovian,davies1976markovian}, and subsequent work of
D\"umcke and Spohn \cite{dumcke1979proper}, establish a positivity-preserving quantum dynamical semigroup in the van Hove limit
$\lambda\rightarrow0$, $t\rightarrow\infty$, with $\lambda^2t$ held
fixed. Nothing in what follows is in tension with these results: our
interest is instead in finite-time maps generated by perturbative,
generally non-Markovian dynamics.

The semigroup requirement (\ref{semigroup}) can be relaxed by allowing a time-dependent generator $\mathcal L(t)$. If $\mathcal L(t)$ is of GKLS form at each
instant, with a positive-semidefinite instantaneous Kossakowski matrix $h(t)$,
then the corresponding evolution is CP-divisible: every {\it intermediate}
propagator $\Phi(t_2,t_1)$ is CPTP. Following \cite{Rivas:2010pzm}, CP-divisibility has become a standard modern notion of Markovianity for time-dependent open-system dynamics, generalizing the more restrictive semigroup notion.

For our purposes, however, the crucial point is that both semigroup
structure and CP-divisibility are stronger requirements than complete
positivity of the finite-time map $\Phi(t,t_{\mathrm{i}})$. Such a map may fail to be CP-divisible (and its instantaneous time-local generator may therefore fail to have positive rates) while the map from the initial time $t_{\mathrm i}$ to every finite time $t$ remains perfectly CPTP. Thus the complete positivity of a finite-time reduced map and the validity of a Markovian master equation are logically distinct questions, as comparisons with exact dynamics have long emphasized \cite{Rivas:2010lbs}.

This distinction is particularly important for non-Markovian open
systems, whose reduced dynamics need not form a semigroup or be CP-divisible
\cite{Maniscalco:2006cfz,Vacchini:2013cug,Hall:2014oiy,
Breuer:2015zlm,Suarez:2025bhb}. Exact non-Markovian master equations,
such as the Hu-Paz-Zhang equation for quantum Brownian motion
\cite{hu1992quantum,halliwell1996alternative}, provide familiar examples
of sensible reduced dynamics lying outside the simplest GKLS setting.
Additional subtleties concerning complete positivity may also arise
from assumptions about the initial system-environment state
\cite{pechukas1994reduced}.

In summary, Redfield-TDC evolution is generically neither a quantum dynamical
semigroup nor CP-divisible. This alone, however, does not determine
whether its finite-time map is CPTP. It is this weaker question that we investigate below.

\section{Complete positivity in perturbation theory}
\label{sec:PT}

We begin in this section by reconsidering weak-coupling perturbation theory at the level of quantum maps. We show that the second-order perturbative correction naturally admits an approximate Kraus representation.

Concretely, we assume the Schr\"odinger-picture Hamiltonian
\begin{equation}
H_{\mathrm{world}}(t)= H_{\mathrm{S}}(t) \otimes \mathbb{I}_{\mathrm{E}} +  \mathbb{I}_{\mathrm{S}} \otimes H_{\mathrm{E}}(t) + \lambda H_{\mathrm{int}}(t),
\end{equation}
where we allow for generic time-dependence of the operators. Here $H_{\mathrm{S}}(t)$ is the system Hamiltonian and $H_{\mathrm{E}}(t)$ the environment Hamiltonian. 
The full Schr\"odinger-picture state $\widetilde{\rho}$ evolves according to the von Neumann equation
\begin{equation}
\frac{\exd \widetilde{\rho}}{\exd t} =  - i [ H_{\mathrm{world}}(t), \widetilde{\rho}(t) ],
\end{equation}
given an initial condition at time $t=t_{\mathrm{i}}$, which we assume is uncorrelated such that
\begin{equation}
\widetilde{\rho}(t_{\mathrm{i}}) = \varrho_{\mathrm{Si}} \otimes \varrho_{\mathrm{Ei}} \; . \label{incond}
\end{equation}
We next define the interaction-picture state and operators in the standard way:
\begin{eqnarray}
\rho(t)
& = &
\mathcal{U}_0^\dagger(t,t_{\mathrm{i}})
\widetilde{\rho}(t)
\mathcal{U}_0(t,t_{\mathrm{i}}) \; , \\
H^{I}_{\mathrm{int}}(t)
& = &
\mathcal{U}_0^\dagger(t,t_{\mathrm{i}})
H_{\mathrm{int}}(t)
\mathcal{U}_0(t,t_{\mathrm{i}})\ , \label{Hint_intP}
\end{eqnarray}
where the free unitary evolution operator is
\begin{equation}
\mathcal{U}_0(t,t_{\mathrm{i}})
=
\mathcal{T}e^{ - i \int_{t_{\mathrm{i}}}^t \mathrm{d} s \; H_{\mathrm{S}}(s) }
\otimes
\mathcal{T}e^{ - i \int_{t_{\mathrm{i}}}^t \mathrm{d} s \; H_{\mathrm{E}}(s) } \; .
\end{equation}
The evolution of the interaction-picture density matrix is then given by
\begin{equation}
\frac{\exd \rho}{\exd t}
=
-i \lambda\, 
\big[
H^{I}_{\mathrm{int}}(t),\rho(t)
\big] \ .
\end{equation}
We can then integrate the above and perturb in the small coupling $\lambda$:
\begin{eqnarray}
\rho(t)
& \simeq &
\varrho_{\mathrm{Si}} \otimes \varrho_{\mathrm{Ei}} \label{perturbation_theory} \\
&& \  -i\lambda
\int_{t_{\mathrm{i}}}^{t}\exd s\,
\left[
H^{I}_{\mathrm{int}}(s), \varrho_{\mathrm{Si}} \otimes \varrho_{\mathrm{Ei}}  
\right] \notag \\
& & \ -
\lambda^2
\int_{t_{\mathrm{i}}}^{t}\exd s
\int_{t_{\mathrm{i}}}^{s}\exd u\, 
\Big[
H^{I}_{\mathrm{int}}(s),
\big[
H^{I}_{\mathrm{int}}(u),\varrho_{\mathrm{Si}} \otimes \varrho_{\mathrm{Ei}} 
\big] \Big] \notag \\
&& \  +\mathcal{O}(\lambda^3) \  . \notag
\end{eqnarray}
We then perform a partial trace over the environment, defining the reduced density matrix for the system
\begin{equation} \label{reduced_DEF}
\varrho_{\mathrm{S}}(t) = \trB \big( \rho(t) \big) \; . 
\end{equation}
With the definition
\begin{equation}
\mathcal{U}(t,t_{\mathrm{i}}) = \mathcal{T}e^{ - i \lambda \int_{t_{\mathrm{i}}}^t \mathrm{d} s \; H^{I}_{\mathrm{int}}(s) } \ ,
\end{equation}
we have
\begin{equation} \label{reduced_Stinespring}
\varrho_{\mathrm{S}}(t) = \trB \big( \mathcal{U}(t,t_{\mathrm{i}}) \, ( \varrho_{\mathrm{Si}} \otimes \varrho_{\mathrm{Ei}}  ) \, \mathcal{U}^{\dagger}(t,t_{\mathrm{i}}) \big) \; . 
\end{equation}
Eq.~(\ref{reduced_Stinespring}) is a Stinespring dilation and therefore defines an exact CPTP map $\Phi(t,t_{\mathrm{i}})$ which admits a Kraus representation \cite{stinespring1955positive,kraus1971general,
alicki2007quantum,rivas2012open,Nielsen:2012yss,Plenio:2014nyj}. Equivalently, by Choi's theorem \cite{choi1975completely}, complete positivity is characterized
by positivity of the corresponding Choi matrix. We write the exact reduced evolution as
\begin{equation}
\varrho_{\mathrm{S}}(t)
=
\Phi(t,t_{\mathrm{i}})
\big[
\varrho_{\mathrm{Si}}
\big] \; .
\label{exactPhi_t_ti}
\end{equation}
Our interest is in how much of this CPTP structure survives when the
exact map $\Phi(t,t_{\mathrm{i}})$ is replaced by its perturbative
approximation.

We take the interaction-picture operator (\ref{Hint_intP}) to take the explicit form
\begin{equation} \label{Hint_SE}
H^{I}_{\mathrm{int}}(t)
=
\sum_{\alpha} S_{\alpha}(t) \otimes E_{\alpha}(t)
\end{equation}
as stated earlier in Eq.~(\ref{Hint_introduction}). Without loss of generality\footnote{This involves no loss of generality. Any non-Hermitian operators appearing in the interaction may be decomposed into their Hermitian and anti-Hermitian parts; Hermiticity of $H^{I}_{\mathrm{int}}(t)$ then ensures that the interaction can be rewritten entirely in terms of Hermitian operators as in Eq.~(\ref{HermitianChoice}).} we take each individual system and environment operator in the above to be Hermitian
\begin{equation} \label{HermitianChoice}
S_{\alpha}^{\dagger}(t) = S_{\alpha}(t) \qquad \mathrm{and} \qquad E_{\alpha}^{\dagger}(t) = E_{\alpha}(t) \; .
\end{equation}
For simplicity, throughout the main text we assume that the environmental operators are centered,
\begin{equation} \label{MFvanish}
\overline E_\alpha(t)
:=
\mathrm{Tr}_{\mathrm E}
\!\left[
E_\alpha(t)\rho_{\mathrm{Ei}}
\right]
=
0\; .
\end{equation}
This assumption is made only to simplify the presentation. When the one-point functions (or mean-field terms) do not vanish, they generate an effective first-order Hamiltonian acting on the system. Passing to the corresponding interaction picture removes these mean-field contributions, after which all of the arguments below proceed unchanged with the centered environmental correlator. For completeness, the general uncentered case is summarized in Appendix~\ref{App:MF}.

Assuming an initially-factorized state (\ref{incond}) and vanishing of mean-field terms (\ref{MFvanish}), taking the partial trace over Eq.~(\ref{perturbation_theory}) gives
\begin{equation}
\rho_{\mathrm S}(t)
=
\rho_{\mathrm{Si}}
+
\lambda^2
\Phi_2(t,t_{\mathrm{i}})
[\rho_{\mathrm{Si}}]
+
\mathcal O(\lambda^3) \; ,
\label{PTmap}
\end{equation}
where a straightforward calculation reveals
\begin{eqnarray}
&& \Phi_2(t,t_\mathrm{i})[\varrho_{\mathrm{Si}}] \label{Phi2_explicit} \\
&& \hspace{5mm} = \sum_{\alpha,\beta}
\int_{t_{\mathrm i}}^{t}\exd s
\int_{t_{\mathrm i}}^{s} \exd u \; 
\bigg( \,  [S_\alpha(u)\varrho_{\mathrm{Si}},S_\beta(s)]
\, \mathcal E_{\beta\alpha}(s,u)
\notag \\
&& \hspace{35mm} 
+ [S_\beta(s),\varrho_{\mathrm{Si}} S_\alpha(u)] \, 
\mathcal E_{\beta\alpha}^{*}(s,u) \, \bigg)\; , \notag 
\end{eqnarray}
and the environmental correlator is $\mathcal E_{\alpha\beta}(t,s)$, defined earlier in Eq.~(\ref{plaindefEenv_introduction}). One can now study properties of the perturbative map itself:,
\begin{equation}
\Phi(t,t_\mathrm{i})
\simeq
\mathcal{I}
+
\lambda^2
\Phi_2(t) + \mathcal{O}(\lambda^3) \ , \label{Phi_PT_IPhi_2}
\end{equation}
where $\mathcal{I}$ is the superoperator identity. Although this expression does not manifestly resemble a CPTP map, the microscopic origin of $\Phi_2$ imposes an important positivity condition on the environmental correlator. The positivity of two-argument correlation functions has been emphasized particularly in the literature on non-Markovian open-system dynamics \cite{Strunz:1996tu,Diosi:1998px,Fleming:2010zqa,Fleming:2011cdb,Diosi:2014upa,Ferialdi:2016agf,Sukharnikov:2025nxm}.

In particular, $\mathcal{E}_{\alpha\beta}(t,s)$ is positive semi-definite as a two-argument kernel: for any collection of test functions $\{g_\alpha(t)\}$, the microscopic correlation function satisfies
\begin{equation}
\sum_{\alpha,\beta}
\int_{t_{\mathrm{i}}}^{t} \exd t'
\int_{t_{\mathrm{i}}}^{t} \exd t''\,
g_\alpha(t')\,
\mathcal{E}_{\alpha\beta}(t',t'')\,
g_\beta^{\ast}(t'')
\geq 0 .
\label{envCORR_positive}
\end{equation}
One denotes this positive-semidefinite two-argument kernel property by $\mathcal{E}_{\alpha\beta}(t,s)\succeq 0$. It follows directly from the microscopic definition of $\mathcal{E}(t,s)$ as a quantum correlation function\footnote{This is an extremely general property of correlation functions (rather than a special assumption about the environment). Indeed, inserting  $O=\sum_\alpha\int_{t_{\mathrm{i}}}^{t}\exd s\, f_\alpha(s)E_\alpha(s)$ into the definition of the correlator in Eq.~(\ref{plaindefEenv_introduction}), one finds $\mathrm{Tr}_{\mathrm E}[O O^{\dagger}\,\varrho_{\mathrm{Ei}}]\geq0$ which reduces to Eq.~(\ref{envCORR_positive}). The argument uses only positivity of the quantum state and so applies equally to finite- and infinite-dimensional environments, including quantum field theories.}.

For continuous kernels on the finite time interval considered here, Mercer's theorem \cite{Mercer:1909} then says that Eq.~(\ref{envCORR_positive}) is equivalent to the mode decomposition Eq.~(\ref{envCORR_intro}), repeated here,
\begin{equation} \label{KernelDecompPT}
\mathcal E_{\alpha\beta}(t,s) = \sum_\kappa u_{\kappa\alpha}(t) u_{\kappa\beta}^{\ast}(s) \; ,
\end{equation}
where the non-negative eigenvalues of the kernel have been absorbed into the mode functions $u_{\kappa\alpha}(t)$. It is this decomposition that plays a central role in what follows.

In microscopic derivations of GKLS master equations \cite{Bochner1933MonotoneFS,Breuer:2002pc,Albash:2012jaz,
Mozgunov:2020dbe}, one additionally specializes to a stationary environment, for which $\mathcal{E}_{\alpha\beta}(t,s) \to \mathcal{E}_{\alpha\beta}(t-s)$. The two-argument kernel positivity $\mathcal{E}_{\alpha\beta}(t,s)\succeq 0$ then implies by Bochner's theorem \cite{46d709a6-dfc1-3520-8cb0-2e4171631157} that the Fourier-transformed correlation function is positive semi-definite (thereby ensuring positivity of the Kossakowski matrix in GKLS generators (\ref{GKLS_generator})). We stop one step earlier: throughout this work we require only the underlying kernel positivity (\ref{envCORR_positive}). We therefore neither assume time-translation invariance nor require a frequency-space representation of the environmental correlations.

Using Eq.~(\ref{KernelDecompPT}), define the operators
\begin{equation} \label{Vkappa_def}
V_\kappa(t,t_{\mathrm{i}})
=
\sum_\alpha
\int_{t_{\mathrm i}}^{t}
ds\;
u_{\kappa\alpha}^{*}(s)
S_\alpha(s)\; .
\end{equation}
With this definition, a straightforward calculation then shows that the second-order correction (\ref{Phi2_explicit}) may always be written as
\begin{equation}
\Phi_2[\varrho_{\mathrm{Si}}]
=
\sum\nolimits_\kappa
V_\kappa
\varrho_{\mathrm{Si}}
V_\kappa^\dagger
-
B \varrho_{\mathrm{Si}}
-
\varrho_{\mathrm{Si}} B^\dagger \; ,
\label{Phi2_B}
\end{equation}
where
\begin{equation}
B(t,t_{\mathrm{i}})
=
\sum_{\alpha,\beta}
\int_{t_{\mathrm i}}^{t}
ds
\int_{t_{\mathrm i}}^{s}
du\;
\mathcal E_{\beta\alpha}(s,u)
S_\beta(s) 
S_\alpha(u) \ . \label{Bdef_main}
\end{equation}
Eq.~(\ref{Phi2_B}) is the central observation of this section.  Further define
\begin{equation}
K_0
=
\mathbb I_{\mathrm{S}}
-
\lambda^2
B 
\ \  \mathrm{and} \ \  
K_\kappa
=
\lambda
V_\kappa  \label{KrusOps_pert}
\end{equation}
where $\mathbb I_{\mathrm{S}}$ is the identity on the system Hilbert space. Using Eq.~(\ref{Phi2_B}), these operators define the completely positive map
\begin{equation}
K_0\varrho_{\mathrm{Si}}K_0^\dagger
+ \sum\nolimits_\kappa
K_\kappa\varrho_{\mathrm{Si}}K_\kappa^\dagger
=
\varrho_{\mathrm{Si}}
+
\lambda^2\Phi_2[\varrho_{\mathrm{Si}}]
+
\mathcal O(\lambda^4).
\label{PerturbativeKraus_pre}
\end{equation}
Since the microscopic perturbative expansion itself is known here only through second order, this agrees with the exact reduced dynamics up to corrections of $\mathcal O(\lambda^3)$, and so 
\begin{equation}
\varrho_{\mathrm{S}}(t) \simeq K_0\varrho_{\mathrm{Si}}K_0^\dagger
+ \sum\nolimits_\kappa
K_\kappa\varrho_{\mathrm{Si}}K_\kappa^\dagger + \mathcal{O}(\lambda^3) \ . \label{PerturbativeKraus}
\end{equation}
Furthermore, upon using $\sum_\kappa V_\kappa^\dagger V_\kappa = B+B^\dagger$ we see perturbative trace preservation:
\begin{equation} \label{TP_pert}
K_0^\dagger K_0 + \sum\nolimits_\kappa
K_\kappa^\dagger
K_\kappa
=
\mathbb{I}_{\mathrm{S}} + \mathcal{O}(\lambda^4) \ .
\end{equation}
Hence ordinary perturbation theory already admits an approximate Kraus representation through second-order in the coupling. The Kraus form makes complete positivity manifest, while Eq.~(\ref{TP_pert}) ensures trace preservation to the same order.

The elephant in the room is that this all remains perturbative. In particular, the operator $\lambda^2 B(t)$ may exhibit secular growth \cite{Fleming:2011irq,Kaplanek:2019dqu,Burgess:2022rdo,Lampert:2025cus}, becoming large at late enough times irrespective of how small the coupling $\lambda $ is. A natural criterion for perturbative control is then
\begin{equation}
\lambda^2 \, | B(t,t_{\mathrm{i}}) | \ll 1\ . 
\end{equation}
Once this condition fails, the approximate Kraus representation has been pushed beyond its regime of validity. This motivates our CPTP resummation of the second-order map in the next section.

\section{A completely positive resolvent resummation}
\label{Sec:Resolvent}

The positive mode decomposition (\ref{KernelDecompPT}) of the environment correlator has a second useful consequence: the perturbative correction $\Phi_2$ can be written in GKLS form. This observation allows us to construct simple resummations of the second-order map which are exactly CPTP.

To see this, define the Hermitian operator 
\begin{equation}
H_B(t,t_{\mathrm{i}})
:=
\frac{B(t,t_{\mathrm{i}})-B^\dagger(t,t_{\mathrm{i}})}{2i} \ .
\label{HBdefinition}
\end{equation}
Using $\sum_\kappa V_\kappa^\dagger V_\kappa = B+B^\dagger$, the perturbative correction (\ref{Phi2_B}) can then be written as
\begin{equation} \label{Phi2GKLSform} 
\Phi_2[\varrho_{\mathrm{Si}}]
=
-i[H_B,\varrho_{\mathrm{Si}}] +
\sum_\kappa
\Big(
V_\kappa \varrho_{\mathrm{Si}} V_\kappa^\dagger
-
\tfrac{1}{2}
\{
V_\kappa^\dagger V_\kappa ,\varrho
_{\mathrm{Si}}
\}
\Big) 
\end{equation}
which is of the familiar GKLS form given in Eq.~(\ref{GKLS_generator}). In the $\kappa$ basis given by the decomposition in Eq.~(\ref{KernelDecompPT}), the Kossakowski matrix is already diagonal and positive-semidefinite.

This means that for each fixed physical time $t$, the superoperator $\Phi_2(t)$ has precisely the form of a GKLS generator. We emphasize that $\Phi_2(t)$ is not being interpreted here as the generator of physical-time evolution. Rather, holding $t$ fixed, introduce an auxiliary parameter $r\geq 0$ through
\begin{equation}
\frac{\partial \sigma_r}{\partial r}
=
\lambda^2
\Phi_2(t,t_{\mathrm{i}})[\sigma_r]
\qquad \mathrm{with} \qquad
\sigma_{0}
=
\varrho_{\mathrm{Si}} \ .
\label{AuxiliaryEvolution}
\end{equation}
Since $\Phi_2$ is of GKLS form, Eq.~(\ref{AuxiliaryEvolution}) generates a CPTP semigroup in the auxiliary parameter $r$, with
\begin{equation}
\sigma_r
=
e^{r\lambda^2\Phi_2(t,t_{\mathrm{i}})}[\sigma_0] \; ,
\end{equation}
so that $e^{r\lambda^2\Phi_2}$ CPTP for every $r\geq0$ (the special case $r=1$ will be useful in \S\ref{sec:Magnus}). This motivates the resolvent resummation 
\begin{equation}
\Phi_{\mathrm{res}}(t,t_{\mathrm{i}})
=
\int_0^\infty
\mathrm{d}r\;
e^{-r}
e^{r\lambda^2\Phi_2(t,t_{\mathrm{i}})} \; .
\label{ResolventIntegral}
\end{equation}
Since $e^{-r}\geq0$ and $\int_0^\infty \mathrm{d}r\,e^{-r}=1$, Eq.~(\ref{ResolventIntegral}) is a positively-weighted average of the CPTP maps $e^{r\lambda^2\Phi_2}$, and is therefore itself CPTP
\cite{Peng:2025apb}. Moreover,
\begin{equation}
\frac{\mathrm{d}}{\mathrm{d}r}
\left(
e^{-r}e^{r\lambda^2\Phi_2}
\right)
=
-
\left(
\mathcal I-\lambda^2\Phi_2
\right)
e^{-r}e^{r\lambda^2\Phi_2} \; ,
\end{equation}
where the eigenvalues of the GKLS generator $\lambda^2\Phi_2$ have non-positive real parts, so the boundary term at $r\rightarrow\infty$ vanishes. Integrating over $r$ then gives
\begin{equation}
\Phi_{\mathrm{res}}(t,t_{\mathrm{i}})
:=
\big( \mathcal{I} -\lambda^2\Phi_2(t,t_{\mathrm{i}})
\big)^{-1} \; .
\label{ResolventDefinition}
\end{equation}
This inverse is guaranteed to exist because all eigenvalues $\gamma$ of
the GKLS generator $\lambda^2\Phi_2$ satisfy $\mathrm{Re}[\gamma]\leq0$. This means $\gamma\neq1$ and so $\mathcal I-\lambda^2\Phi_2$ has no zero eigenvalues (and is therefore invertible).

The resolvent (\ref{ResolventDefinition}) also admits a simple self-consistent interpretation. Since Eq.~(\ref{PTmap}) implies $\varrho_{\mathrm{Si}} = \varrho_{\mathrm{S}}(t)+\mathcal{O}(\lambda^2)$, replacing the initial state by the final state inside the second-order correction changes that term only at $\mathcal{O}(\lambda^4)$. This suggests the self-consistency relation
\begin{equation}
\varrho_{\mathrm{S}}(t)
\simeq
\varrho_{\mathrm{Si}}
+
\lambda^2
\Phi_2(t)
\big[
\varrho_{\mathrm{S}}(t)
\big] \; ,
\label{SelfConsistentPT}
\end{equation}
whose solution is
\begin{equation}
\varrho_{\mathrm{S}}(t)
\simeq
\big(
\mathcal I
-
\lambda^2
\Phi_2(t)
\big)^{-1}
\big[
\varrho_{\mathrm{Si}}
\big] \ ,
\end{equation}
giving the resolvent map $\Phi_{\mathrm{res}}(t)$ from Eq.~(\ref{ResolventDefinition}). 

Notice that expanding the resolvent map in the coupling $\lambda$ yields
\begin{equation}
\Phi_{\mathrm{res}}
=
\mathcal I
+
\lambda^2\Phi_2
+
\lambda^4\Phi_2^2
+\cdots
\end{equation}
which agrees with perturbation theory through second-order, while resumming repeated insertions of $\Phi_2$, reminiscent of a Pad\'e or self-energy resummation in quantum field theory.

We do not expect this elementary resummation to reproduce generic late-time dynamics; in particular, it need not capture the exponential relaxation characteristic of many weakly-coupled open systems. Indeed, in Fig.~\ref{Fig:qubitqubit_comparison} we find that the resolvent fails to track the exact dynamics at late times. Its significance here is primarily structural: it provides a very simple CPTP completion of the second-order perturbative map and illustrates explicitly the non-uniqueness of such resummations. It may nevertheless provide a useful approximation in problems where intermediate-time dynamics, rather than the asymptotic late-time behaviour, is of primary interest.

\section{The Redfield equation as a time-local resummation}
\label{sec:Redfield}

We now turn to the time-local resummation that follows most directly
from the microscopic dynamics: the Redfield equation with time-dependent
coefficients, or equivalently the second-order time-convolutionless
(TCL$_2$) master equation. Rather than deriving this equation using
projection-operator methods, we obtain it directly at the level of
quantum maps.

As in Eq.~(\ref{exactPhi_t_ti}), the exact reduced dynamics may be written as
\begin{equation}
\varrho_{\mathrm S}(t)
=
\Phi(t,t_{\mathrm i})
\big[
\varrho_{\mathrm{Si}}
\big] \; ,
\end{equation}
where $\Phi(t,t_{\mathrm i})$ is the exact CPTP evolution map. Provided this map is invertible (necessarily true at sufficiently early times given Eq.~(\ref{PTmap})), the evolution can be written in time-local form. Differentiating with respect to the final time and eliminating the initial state using
\begin{equation}
\varrho_{\mathrm{Si}}
=
\Phi^{-1}(t,t_{\mathrm i})
\big[
\varrho_{\mathrm S}(t)
\big]
\end{equation}
gives the exact identity \cite{Hall:2014oiy}
\begin{equation}
\frac{\mathrm d \varrho_{\mathrm S}}{\mathrm dt}
=
\dot{\Phi}(t,t_{\mathrm i})
\Phi^{-1}(t,t_{\mathrm i})
\big[
\varrho_{\mathrm S}(t)
\big] \; .
\label{ExactTCL}
\end{equation}
Thus the exact time-local generator is $\dot{\Phi}\Phi^{-1}$ whenever the inverse of the exact map exists.

We now truncate this generator perturbatively. From Eq.~(\ref{Phi_PT_IPhi_2}) one finds:
\begin{align}
\Phi^{-1}(t,t_{\mathrm i})
&=
\mathcal I
-
\lambda^2\Phi_2(t,t_{\mathrm i})
+
\mathcal O(\lambda^3),
\\
\dot{\Phi}(t,t_{\mathrm i})
&=
\lambda^2\dot{\Phi}_2(t,t_{\mathrm i})
+
\mathcal O(\lambda^3),
\end{align}
This means that to second-order,
\begin{equation}
\frac{\mathrm d \varrho_{\mathrm S}}{\mathrm dt}
\simeq 
\lambda^2
\dot{\Phi}_2(t,t_{\mathrm i})
\big[
\varrho_{\mathrm S}(t)
\big] .
\label{RedfieldEq_intermsof_Phi2}
\end{equation}
as stated in Eq.~(\ref{IntroRedfield}), which is precisely the Redfield TDC. Using the expression (\ref{Phi2_explicit}) one can write this in the more explicit form presented in Eq.~(\ref{Red_H_centered_introduction}) in the introduction.

The derivation of the Redfield equation from microscopic system-environment dynamics is standard textbook material
(see {\it eg.}~\cite{Breuer:2002pc}). Historically, it was introduced by Redfield in 1957 in the context of spin relaxation theory \cite{redfield1957theory,redfield1965theory}, nearly two decades before the GKLS theorem\footnote{An equivalent master equation (see Ref.~\cite{timm2008tunneling}) was derived by Bloch and Wangsness in 1953 \cite{wangsness1953dynamical}.}. Higher orders in the expansion of $\dot{\Phi}\Phi^{-1}$ give the corresponding higher-order TCL generators.

The same equation can be obtained in several equivalent ways. The standard Born approximation replaces the evolving system-environment state by an appropriately factorized state, while the time-convolutionless projector expansion
\cite{shibata1977generalized,chaturvedi1979time,shibata1980expansion} directly generates the TCL$_2$ equation when the reduced map is invertible; recent work extends this using a pseudoinverse when ordinary invertibility fails \cite{Blumenfeld:2025lkn}. By contrast, the Nakajima-Zwanzig equation \cite{Nakajima:1958pnl,Zwanzig:1960gvu} retains an explicit memory integral over the earlier reduced state and is generally more difficult to solve \cite{kleinekathofer2004non,schroder2007reduced,Moustos:2016lol}. Direct comparisons have nevertheless found TCL approximations to be more accurate than their Nakajima-Zwanzig counterparts in some models \cite{Shen:2014gcv}. Our concern here is not to adjudicate among these master equations, but rather to study the finite-time map generated by Eq.~(\ref{RedfieldEq_intermsof_Phi2}).

Indeed, Eq.~(\ref{RedfieldEq_intermsof_Phi2}) should be distinguished from the perturbative map $\Phi_2$ from Eq.~(\ref{PTmap}). Solving the second-order time-local equation generates contributions to all orders in $\lambda$. Formally, its solution is
\begin{equation}
\Phi_{\mathrm{Red}}(t,t_{\mathrm i})
=
\mathcal T
\exp
\left[
\lambda^2
\int_{t_{\mathrm i}}^{t}
\mathrm ds\,
\dot{\Phi}_2(s,t_{\mathrm i})
\right],
\label{RedfieldMap}
\end{equation}
where $\mathcal T$ denotes superoperator time-ordering (see Appendix \ref{App:Magnus}). Expanding Eq.~(\ref{RedfieldMap}) gives $\mathcal I+\lambda^2\Phi_2 + \ldots$ through second-order, but the time-ordered exponential supplies a particular infinite set of higher-order terms. Redfield evolution is therefore a specific resummation of the same second-order perturbative $\Phi_2$ considered in the preceding sections.

Unlike the resolvent map of \S\ref{Sec:Resolvent}, however, Eq.~(\ref{RedfieldMap}) is {\it not} manifestly CPTP. In particular, the instantaneous generator $\dot{\Phi}_2(t)$ is generically not of GKLS form. It can be rearranged into a Lindblad-{\it like} form\footnote{Interestingly this suggests the Redfield equation may be phrased as gradient flow along the lines of \cite{Kaplanek:2025pzi}.} with time-dependent coefficients\footnote{See, for example, Appendix A of Ref.~\cite{Ferialdi:2017dnt}, which performs this rearrangement for a system with a quadratic Hamiltonian, where one may expand $S_{\alpha}(s)=\sum_{\beta}U_{\alpha\beta}(s-t)S_{\beta}(t)$.}, where some instantaneous rates may be negative as seen {\it eg.}~in Eq.~(\ref{eq:qubitRedfield_GKLSform}). As emphasized in \S\ref{Sec:Lindblad} however, a failure of the instantaneous generator to be GKLS implies failure of CP divisibility, {\it not} necessarily failure of complete positivity of the finite-time map.

Concerns about Redfield evolution nevertheless have a long history. Positivity violations\footnote{For certain classes of qubit dynamics, positivity is sufficient
to guarantee complete positivity \cite{Hall:2008jez}.} have been identified in a variety of models \cite{suarez1992memory,gaspard1999slippage,Benatti:2005uq, zhao2002density,Jeske:2015mlq,Benatti:2022ect,delaPradilla:2024cnp,Tokieda:2023yig,Tupkary:2023txe}, and are routinely discussed in textbooks \cite{van1992stochastic,breuerTheoryOpenQuantum2002,alicki2007quantum}. A standard remedy is an additional rotating-wave approximation, although this may discard physically relevant information \cite{alicki1989master}. A substantial literature has therefore developed alternative regularizations, coarse-grainings, and positivity-preserving modifications of weak-coupling master equations \cite{Schaller:2008jiu,Schaller:2009jxb,Majenz:2013piv,
kirvsanskas2018phenomenological,Farina:2019fcd,McCauley:2020dre,
Mozgunov:2020dbe,Nathan:2020qck,Chen:2020mta,Davidovic:2020wdg,
Potts:2021kvr,Becker:2021xcj,Trushechkin:2021vbr,Davidovic:2021mlg,
DAbbruzzo:2022hgp,DAbbruzzo:2023sds,Nicacio:2024xxf}; see also the recent overview in Ref.~\cite{DAbbruzzo:2026ibj}. Related approaches reorganize the weak-coupling expansion itself to obtain better-controlled approximations \cite{rivas2017refined,Tupkary:2021bcd,Lee:2022wvh}.

There is, however, substantial evidence that finite-time Redfield evolution can behave considerably better than this standard picture might suggest. Early studies found that retaining the time dependence of the coefficients can substantially improve the evolution \cite{Yu:1999yy,yu2000post,Maniscalco:2003hck,cheng2005markovian,Whitney:2008lcj}. This should be distinguished from the additional late-time approximation in which the upper limit of the memory integral is extended to infinity, which can introduce positivity violations, initial-condition ``slippage'', and poor short-time behaviour \cite{suarez1992memory,gaspard1999slippage,zhao2002density,yu2000post,Whitney:2008lcj}. We therefore work with the finite-time Redfield equation (\ref{RedfieldEq_intermsof_Phi2}), which follows directly from the perturbative map $\Phi_2(t,t_{\mathrm i})$, without making the additional
asymptotic approximation of extending the memory integral to infinity. Studies supporting the practical use of the latter despite its positivity problems include Refs.~\cite{Jeske:2015mlq,Eastham:2016vai}.

More recent results reinforce that the Redfield TDC is better behaved than has traditionally been assumed. Certain weak-coupling limits of Redfield TDC can agree remarkably well with exact dynamics in driven and non-Markovian models \cite{Shen:2014gcv,jirauschek2019optoelectronic,link2022non}, while Ref.~\cite{Hartmann:2019rpc} finds that small positivity violations occur predominantly as the weak-coupling approximation itself begins to fail. Ref.~\cite{Tokieda:2024soc} provides an explicit example in which Redfield TDC evolution remains completely positive for all times, and Ref.~\cite{Tokieda:2025kcr} similarly argues that positivity should be expected within an appropriate weak-coupling regime. Conversely, positivity violations of time-dependent Redfield evolution can occur at sufficiently strong coupling \cite{Litzba:2024cwd}.

These results suggest that the relevant question is not simply whether the instantaneous Redfield generator is of GKLS form, but whether its finite-time resummation remains CPTP within a controlled weak-coupling regime. Existing results point toward such a regime, but do not provide a general quantitative criterion for it. Our goal in what follows is to make this statement precise: under what conditions does the finite-time Redfield map (\ref{RedfieldMap}) remain close to a CPTP resummation of the second-order perturbative map?

\section{Magnus resummation}
\label{sec:Magnus}

The Redfield equation provides a physically motivated resummation of perturbation theory, but unlike the resolvent map of
\S\ref{Sec:Resolvent}, its complete positivity is not manifest. We now show that the Redfield map is naturally associated with yet another CPTP resummation, with the difference between the two controlled by superoperator time-ordering.

Recall from Eq.~(\ref{Phi2GKLSform}) that, for each fixed physical time $t$, the second-order perturbative map $\Phi_2(t,t_{\mathrm{i}})$ has the form of a GKLS generator. We therefore define the {\it Magnus map}\footnote{Magnus expansions have also been applied directly to Lindblad master equations \cite{Haddadfarshi:2014sdi}, although this is distinct from their use here.}
\begin{equation}
\Phi_{\mathrm{Mag}}(t,t_{\mathrm i})
:=
e^{\lambda^2\Phi_2(t,t_{\mathrm i})}\; .
\label{MagnusMap}
\end{equation}
As in the auxiliary evolution introduced in Eq.~(\ref{AuxiliaryEvolution}), $t$ here labels the GKLS generator rather than the parameter of the associated semigroup. It follows immediately that $\Phi_{\mathrm{Mag}}$ is CPTP. Importantly, despite the CPTP nature of $\Phi_{\mathrm{Mag}}$, it need not form a semigroup or be CP divisible, and hence need not be Markovian in either of the senses discussed in \S\ref{Sec:Lindblad}. Moreover,
\begin{equation}
\Phi_{\mathrm{Mag}}
=
\mathcal I
+
\lambda^2\Phi_2
+
\frac{\lambda^4}{2}\Phi_2^2
+
\mathcal O(\lambda^6),
\end{equation}
so it agrees with the perturbative map through second-order.

The Magnus map (\ref{MagnusMap}) has appeared previously in several forms, including the cumulant equation of Alicki
\cite{alicki1989master}, dynamical coarse-graining \cite{Schaller:2008jiu}, and the refined weak-coupling approach
\cite{rivas2017refined,Rivas:2019wug}; see also ~\cite{Winczewski:2021vvg,Lobejko:2022pab,Suarez:2024vol}. These works assume time-translation-invariant environment correlations and use the positivity of the corresponding Fourier-transformed correlation matrix. Although not stated explicitly, this positivity ultimately follows from Bochner's theorem. Here we instead have worked directly with the positivity of the kernel $\mathcal E_{\alpha\beta}(t,s)$, so neither time-translation invariance nor a Fourier-space description is required.

There are, of course, many CPTP completions of the same perturbative map. For example, the resolvent map of \S\ref{Sec:Resolvent} contains the expansion
\begin{equation}
\Phi_{\mathrm{res}}
=
\mathcal I
+
\lambda^2\Phi_2
+
\lambda^4\Phi_2^2
+\cdots,
\end{equation}
and therefore differs from $\Phi_{\mathrm{Mag}}$ beginning at higher orders. Perturbation theory and complete positivity alone consequently do not single out the exponential map. What distinguishes $\Phi_{\mathrm{Mag}}$ for our purposes is its direct relation to Redfield TDC evolution.

Recall that the Redfield map from Eq.~(\ref{RedfieldMap}) is
\begin{equation}
\Phi_{\mathrm{Red}}(t,t_{\mathrm i})
=
\mathcal T
\exp\left[
\lambda^2
\int_{t_{\mathrm i}}^t
\exd s\,
\dot{\Phi}_2(s,t_{\mathrm i})
\right].
\label{RedfieldMap_Magnus}
\end{equation}
The Magnus expansion (see Appendix \ref{App:Magnus}) makes precise the relation between $\Phi_{\mathrm{Mag}}$ and $\Phi_{\mathrm{Red}}$. Writing the time-ordered exponential in a Magnus expansion gives
\begin{equation}
\Phi_{\mathrm{Red}}(t,t_{\mathrm i})
=
\exp\left[
\lambda^2\Omega_2(t,t_{\mathrm i})
+
\lambda^4\Omega_4(t,t_{\mathrm i})
+
\mathcal O(\lambda^6)
\right],
\label{RedMagnusExpansion}
\end{equation}
the leading term is
\begin{equation}
\Omega_2(t,t_{\mathrm i})
=
\int_{t_{\mathrm i}}^t
\exd s\,
\dot{\Phi}_2(s,t_{\mathrm i})
=
\Phi_2(t,t_{\mathrm i}) \; ,
\label{MagnusOmega1}
\end{equation}
since $\Phi_2(t_{\mathrm i},t_{\mathrm i})=0$.

Meanwhile the first correction due to time ordering is
\begin{equation}
\Omega_4(t,t_{\mathrm i})
=
\frac{1}{2}
\int_{t_{\mathrm i}}^t
\exd s
\int_{t_{\mathrm i}}^s
\exd u\,
\left[
\dot{\Phi}_2(s,t_{\mathrm i}),
\dot{\Phi}_2(u,t_{\mathrm i})
\right].
\label{MagnusOmega2}
\end{equation}
Thus $\Phi_{\mathrm{Mag}}=e^{\lambda^2\Phi_2}$ is precisely the leading Magnus approximation to the Redfield map. The first distinction between the two resummations is entirely due to unequal-time commutators of the Redfield generator. The first Magnus correction can equivalently be written as
\begin{equation}
\Omega_4(t,t_{\mathrm i})
=
\frac{1}{2}
\int_{t_{\mathrm i}}^t
\exd s\,
\left[
\dot{\Phi}_2(s,t_{\mathrm i}),
\Phi_2(s,t_{\mathrm i})
\right].
\label{MagnusOmega2_single}
\end{equation}
In particular, if
\begin{equation}
\left[
\dot{\Phi}_2(s,t_{\mathrm i}),
\dot{\Phi}_2(u,t_{\mathrm i})
\right]
=
0
\qquad
\text{for all }s,u >t_{\mathrm i} \; ,
\label{CommutingRedfield}
\end{equation}
then all higher Magnus corrections vanish and
\begin{equation}
\Phi_{\mathrm{Red}}(t,t_{\mathrm i})
=
\Phi_{\mathrm{Mag}}(t,t_{\mathrm i}) \; .
\end{equation}
The Redfield map is then necessarily CPTP. If (\ref{CommutingRedfield}) fails then one find correspondingly, that the two maps differ at leading order by
\begin{equation}
\Phi_{\mathrm{Red}}(t,t_{\mathrm i})
-
\Phi_{\mathrm{Mag}}(t,t_{\mathrm i})
=
\lambda^4\Omega_4(t,t_{\mathrm{i}})
+
\mathcal O(\lambda^6).
\label{RedMinusMagnus}
\end{equation}
When the higher Magnus corrections are small, the Redfield map is
close to the manifestly CPTP Magnus map. This means that in general the relative size of the first correction $\lambda^4 \Omega_{4}$ relative to the leading-order term $\lambda^2 \Omega_2$ in Eq.~(\ref{RedMagnusExpansion}) provides a simple measure of the importance of time-ordering. We define
\begin{equation}
\epsilon_{\mathrm{Mag}}(t,t_{\mathrm i})
:= 
\frac{\lambda^2}{2} \, 
\frac{
\Big|
\displaystyle
\int_{t_{\mathrm i}}^t
\exd s\,
\left[
\dot{\Phi}_2(s,t_{\mathrm i}),\Phi_2(s,t_{\mathrm i})
\right]
\Big|
}{
\big|\Phi_2(t,t_{\mathrm i})\big|
}.
\label{epMag_def}
\end{equation}
The condition $\epsilon_{\mathrm{Mag}}\ll1$ means that the first time-ordering correction to the Redfield exponent is parametrically small compared with the leading term. If the higher Magnus corrections remain similarly controlled, as suggested by their higher-order dependence on the coupling, then $\Phi_{\mathrm{Red}}$ approaches the manifestly CPTP $\Phi_{\mathrm{Mag}}$. We regard $\epsilon_{\mathrm{Mag}}$ as a simple practical diagnostic rather than a rigorous convergence criterion.

Importantly, $\epsilon_{\mathrm{Mag}}\ll1$ does not by itself prove that $\Phi_{\mathrm{Red}}$ is CPTP. Rather, it isolates the contribution of time ordering to the Redfield exponent and provides a simple quantitative diagnostic for when the Redfield map approaches a guaranteed-CPTP resummation. As we shall see below, this criterion appears to work well in practice in the examples we consider.

For the finite-dimensional examples below we evaluate Eq.~(\ref{epMag_def}) using the Hilbert-Schmidt norm of the matrix representation of the superoperator. For infinite-dimensional systems, where the relevant superoperators may be unbounded, additional care is required in formulating an analogous criterion.

\section{Complete positivity for a qubit: Redfield vs. Magnus maps}
\label{sec:qubitCP}

We now specialize to the simplest nontrivial example in which we can differentiate the Redfield and Magnus maps: a single qubit interacting with an arbitrary environment. Our goal is to understand the structure of the Redfield and Magnus maps, particularly the extent to which they satisfy (or violate) complete positivity.

We consider a qubit with free Hamiltonian
\begin{equation}
H_{\mathrm S}
=
\frac{\omega}{2}\sigma_3,
\end{equation}
coupled linearly to the environment through the interaction $H_{\mathrm{int}} = \sigma_1 \otimes E$. In the interaction picture we have
\begin{equation} \label{generalqubit_Hint}
H_{\mathrm{int}}^{I}(t)
=
\sigma_1(t)\otimes E(t)\ ,
\end{equation}
where
\begin{equation}
\sigma_1(t)
:=
e^{iH_{\mathrm S}t}
\sigma_1
e^{-iH_{\mathrm S}t}
=
\sigma_+e^{i\omega t}
+
\sigma_-e^{-i\omega t}, \
\end{equation}
using the standard raising and lowering operator
\begin{equation}
\sigma_{\pm} := \frac{ \sigma_{1} \pm i \sigma_2 }{2},
\end{equation}
and assuming an initial state at time $t_{\mathrm{i}} =0$ such that 
\begin{equation}
\widetilde{\rho}(0) = \varrho_{\mathrm{Si}} \otimes \varrho_{\mathrm{Ei}}  \ .
\end{equation}
We are agnostic about what the environment is at this point, only assuming (as before) a vanishing environment one-point function,
\begin{equation}
\overline E(t)=0\ .
\end{equation}
The single environment correlator takes the form of Eq.~(\ref{KernelDecompPT}), where  
\begin{eqnarray} \label{KernelQUBIT}
\mathcal E(t,s) = \trB
\big[
E(t)
E(s)
\rho_{\mathrm{Ei}}
\big] = \sum_\kappa
u_{\kappa}(t)
u_{\kappa}^{*}(s) \ .
\end{eqnarray}
If the environment is continuous, we replace the trace and summation by the analogous integral expressions.

\subsection{The Redfield map for a qubit}

The corresponding Redfield equation is ({\it cf.}~Eq.~(\ref{Red_H_centered_introduction}))
\begin{eqnarray}
\frac{\exd \varrho_{\mathrm{S}}}{\exd t} & \simeq & \lambda^2
\int_{0}^{t}\exd s \, 
\bigg( \,  [\sigma_1(s)\varrho_{\mathrm{S}}(t),\sigma_1(t)]
\, \mathcal E(t,s) \label{Red_H_centered_QUBITgen} \\
&& \hspace{14.5mm} 
+ [\sigma_1(t),\varrho_{\mathrm{S}}(t) \sigma_1(s)] \, 
\mathcal E^{*}(t,s) \, \bigg)\ . \notag 
\end{eqnarray}
For convenience we define the family of functions
\begin{equation}
f_{\pm}(t) := \int_{0}^{t} \exd s\; \mathcal{E}(t,s) e^{\pm i \omega (t - s)}, \label{fpm_def} 
\end{equation}
which usefully rewrites the integral as a sum:
\begin{eqnarray}
\int_0^t \exd s \; \mathcal{E}(t,s) \sigma_{1}(s) & = & e^{+ i \omega t} f_{-}(t) \sigma_{+} + e^{- i \omega t} f_{+}(t) \sigma_{-} \notag \\ 
& = & \sum_{\mathfrak{a} = \pm } e^{ \mathfrak{a} i \omega t} f_{-\mathfrak{a}}(t) \sigma_{\mathfrak{a}}.  
\end{eqnarray}
With this substitution Eq.~(\ref{Red_H_centered_QUBITgen}) can be rewritten more simply:
\begin{eqnarray}
\frac{\exd \varrho_{\mathrm{S}}}{\exd t} & \simeq & \lambda^2 \sum_{\mathfrak{a},\mathfrak{b} = \pm} e^{\mathfrak{a}i\omega t} e^{ \mathfrak{b} i  \omega t} \Big(  f_{-\mathfrak{a}}(t) [\sigma_{\mathfrak{a}}\varrho_{\mathrm{S}}(t),\sigma_{\mathfrak{b}}] \qquad  \\
&& \hspace{26mm} + \, f^{\ast}_{\mathfrak{a}}(t) [\sigma_{\mathfrak{b}}, \varrho_{\mathrm{S}}(t) \sigma_{\mathfrak{a}} ] \, \Big) .\notag
\end{eqnarray}
Using $\sigma_{\pm}^{\dagger} = \sigma_{\mp}$ one may manipulate the above expression into a GKLS-{\it like} form ({\it cf.}~Eq.~(\ref{GKLS_eq}))
\begin{eqnarray} \label{eq:qubitRedfield_GKLSform}
\frac{d\varrho_{\mathrm S}}{dt} &
=& - i  \big[ \tfrac{\delta(t)}{2} \sigma_3 ,  \varrho_{\mathrm S}(t) \big] \\
&& + \sum_{\mathfrak{a}, \mathfrak{b} = \pm } h_{\mathfrak{a}\mathfrak{b}}(t) \Big(  \sigma_{\mathfrak{a}} \varrho_{\mathrm S}(t) \sigma^{\dagger}_{\mathfrak{b}} - \tfrac{1}{2} \{ \sigma_{\mathfrak{b}}^{\dagger}  \sigma_{\mathfrak{a}} , \varrho_{\mathrm{S}}(t) \} \Big). \notag
\end{eqnarray}
There is thus a time-dependent shift in the qubit gap,
\begin{equation}
\delta(t) = \lambda^2\mathrm{Im}[ f_{+}(t) - f_{-}(t) ] \ ,
\end{equation}
as well as a Kossakowski matrix of the form
\begin{equation}
[ h_{\mathfrak{a}\mathfrak{b}}(t ) ]  = \lambda^2 \left[ \begin{smallmatrix}  2 \mathrm{Re}[f_{-}(t)] & e^{2 i \omega t}[ f_{-}(t) + f_{+}^{\ast}(t) ] \\
e^{-2 i \omega t}[ f^{\ast}_{-}(t) + f_{+}(t) ] & 2 \mathrm{Re}[f_{+}(t)]  \end{smallmatrix} \right] \ .
\end{equation}
Eq.~(\ref{eq:qubitRedfield_GKLSform}) is manifestly trace-preserving, as well as Hermiticity-preserving since $h_{\mathfrak{a}\mathfrak{b}}(t)$ is Hermitian. However, the coefficients above are generically time-dependent, and therefore the evolution does not\footnote{If one insists on GKLS evolution, then at times much later than the decay time of $\mathcal{E}$ one may sometimes approximate $f_{\pm}(t)$ as constants. An additional rotating-wave approximation removes the remaining oscillatory terms $e^{\pm 2i\omega t}$. We do not pursue these additional approximations here.} generally form a quantum dynamical semigroup in the original sense of GKLS \cite{Lindblad:1975ef,Gorini:1976cm}. We can also ask when the evolution is CP-divisible \cite{Alicki:1979wxy,davies1978open,Breuer:2004cdm,Breuer:2009pku,Rivas:2010pzm,Plenio:2014nyj}, which requires the instantaneous Kossakowski matrix $h(t)$ to be positive-semidefinite. Its eigenvalues are
\begin{eqnarray}
\uplambda^{(h)}_{\pm}(t) & = & \lambda^2 \mathrm{Re}[ f_{-}(t) + f_{+}(t) ] \\
&& \pm \lambda^2 \sqrt{ | f_{-}(t) + f_{+}^{\ast}(t) |^2 + \mathrm{Re}[ f_{-}(t) - f_{+}(t) ]^2 } \ . \notag
\end{eqnarray}
Hence requiring $h(t)$ to be positive semidefinite implies
\begin{equation}
\uplambda^{(h)}_{\pm}(t) \geq 0 \ \  \implies \ \ f_{+}(t) = f_{-}(t)\ \& \ \mathrm{Re}[f_{\pm}(t)] \geq 0 \ . 
\end{equation}
Examining the determinant,
\begin{equation}
\det h(t) = - \lambda^4 | f_{-}(t) - f_{+}(t) |^2 \leq 0 \ , 
\end{equation}
shows that $h(t)$ necessarily has one negative eigenvalue whenever $f_{+}(t) \neq f_{-}(t)$. Generic Redfield evolution is therefore neither a quantum dynamical semigroup nor CP-divisible. As we shall show, this does not preclude the finite-time evolution from being well defined and CPTP: in the regime where the Redfield map is well-approximated by the Magnus map, complete positivity can nevertheless be maintained. 

To this end, let us explore the structure of the Redfield map further. Adopting the notation of Eq.~(\ref{RedfieldEq_intermsof_Phi2}), the Redfield equation is here
\begin{equation}
\frac{\mathrm d \varrho_{\mathrm{S}} }{\mathrm d t}
\simeq
\lambda^2
\dot{\Phi}_2(t)
\big[
\varrho_{\mathrm S}(t)
\big].  \label{RedfieldEq_intermsof_Phi2_QUBIT}
\end{equation}
Explicitly ordering the reduced density matrix $\varrho_{\mathrm{S}}$ as
\begin{equation} \label{explicitordering}
\varrho_{\mathrm{S}}(t) = \left[ \begin{matrix} \varrho_{00}(t) & \varrho_{01}(t) \\
\varrho_{10}(t) & \varrho_{11}(t) \end{matrix} \right] \; \to \; \left[ \begin{matrix} \varrho_{00}(t) \\ \varrho_{11}(t) \\ \varrho_{01}(t) \\
\varrho_{10}(t) \end{matrix} \right] \ ,
\end{equation}
we see Eq.~(\ref{eq:qubitRedfield_GKLSform}) is the matrix
\begin{equation}
\dot{\Phi}_2(t) = \left[
\begin{smallmatrix}
-2\mathrm{Re}f_{+} & 2\mathrm{Re}f_{-} & 0 & 0 \\ 
2\mathrm{Re}f_{+} & - 2\mathrm{Re}f_{-} & 0 & 0 \\
0 & 0 & - f_{-} - f_{+}^{\ast} & e^{2 i \omega t}[ f_{-} + f_{+}^{\ast} ] \\
0 & 0 & e^{-2 i \omega t}[ f^{\ast}_{-} + f_{+} ] & - f_{-}^{\ast} - f_{+} 
\end{smallmatrix}
\right] \ . \label{Phi2_supermatrix}
\end{equation}
The Redfield generator is generically not of GKLS form, but this does not by itself determine whether the finite-time Redfield map is completely positive. We therefore turn directly to the Redfield map Eq.~(\ref{RedfieldMap}):
\begin{equation}
\Phi_{\mathrm{Red}}(t)
=\mathcal T
\exp
\left[
\lambda^2
\int_{0}^{t}
\exd s \; 
\dot{\Phi}_2(s)
\right] \ .
\label{RedfieldMap_qubit}
\end{equation}
Although the time-ordered exponential cannot generally be evaluated in closed form, its structure follows immediately from the form of $\dot{\Phi}_2(t)$. Since Eq.~(\ref{Phi2_supermatrix}) is block-diagonal in the population and coherence sectors, the same is true of $\Phi_{\mathrm{Red}}(t)$. Moreover, trace and Hermiticity preservation\footnote{Time-exponentiating Eq.~(\ref{Phi2_supermatrix}) preserves its block-diagonal structure, so that $\Phi_{\mathrm{Red}}(t)=\mathrm{diag}(P,Q)$, where $P$ and $Q$ are $2\times2$ matrices acting on the population and coherence sectors, respectively. With the operator ordering introduced in Eq.~(\ref{explicitordering}), trace preservation $\mathrm{Tr}(\Phi_{\mathrm{Red}}(t)[X])=\mathrm{Tr}(X)$ implies $\sum_{i} P_{ij} = 1$ for each $j$, while Hermiticity preservation $\Phi_{\mathrm{Red}}(t)[X^{\dagger}]=\Phi_{\mathrm{Red}}(t)[X]^{\dagger}$ implies $P^{\ast}=P$ and $\sigma_1 Q^{\ast}\sigma_1=Q$. Together, these conditions fix the stated form in Eq.~(\ref{PhiRed_BLOCKqubit})} fix the structure to be
\begin{equation}
\Phi_{\mathrm{Red}}(t) = \left[
\begin{matrix}
a_{\mathrm{Red}}(t) & b_{\mathrm{Red}}t) & 0 & 0 \\ 
1-a_{\mathrm{Red}}(t) & 1-b_{\mathrm{Red}}(t) & 0 & 0 \\
0 & 0 & c_{\mathrm{Red}}(t) & d_{\mathrm{Red}}^{\ast}(t) \\
0 & 0 & d_{\mathrm{Red}}(t) & c_{\mathrm{Red}}^{\ast}(t)
\end{matrix}
\right], \  \label{PhiRed_BLOCKqubit}
\end{equation}
for two real functions $a_{\mathrm{Red}}(t)$, $b_{\mathrm{Red}}(t)$ and two complex functions $c_{\mathrm{Red}}(t)$, $d_{\mathrm{Red}}(t)$. Since the map $\Phi_{\mathrm{Red}}$ reduces to the identity at $t=0$, the initial conditions are
\begin{equation} \label{abcd_IC}
a_{\mathrm{Red}}(0)=1\; ,\quad
b_{\mathrm{Red}}(0)=0\; ,\quad
c_{\mathrm{Red}}(0)=1\; ,\quad
d_{\mathrm{Red}}(0)=0 \; .
\end{equation}
Complete positivity can now be tested directly at the level of the finite-time map. By the Choi theorem, $\Phi_{\mathrm{Red}}(t)$ is completely positive if and only if
\begin{equation}
C_{\mathrm{Red}}(t)
=
\sum_{i,j=0}^{1}
|i\rangle\langle j|
\otimes
\Phi_{\mathrm{Red}}(t)
[|i\rangle\langle j|]
\end{equation}
is positive semidefinite. The Choi matrix is a reshuffling of the above:
\begin{equation}
C_{\mathrm{Red}}
=
\begin{pmatrix}
a_{\mathrm{Red}} & 0 & 0 & c_{\mathrm{Red}}
\\
0 & 1-a_{\mathrm{Red}} & d_{\mathrm{Red}} & 0
\\
0 & d_{\mathrm{Red}}^* & b_{\mathrm{Red}} & 0
\\
c_{\mathrm{Red}}^* & 0 & 0 & 1-b_{\mathrm{Red}}
\end{pmatrix}.
\label{eq:qubitChoi}
\end{equation}
Notice that this matrix is Hermitian (unlike $\Phi_{\mathrm{Red}}$). The eigenvalues of the Choi matrix are 
\begin{small}
\begin{equation}
\begin{aligned}
\label{CRed_eigs}
\uplambda^{(C_{\mathrm{Red}})} & \in \Big\{ \tfrac{ ( 1+a_{\mathrm{Red}}-b_{\mathrm{Red}} ) \, \pm \sqrt{ (1-a_{\mathrm{Red}}-b_{\mathrm{Red}})^2 + 4 |c_{\mathrm{Red}}|^2 } }{2} \ , \\
& \hspace{6mm} \tfrac{(1-a_{\mathrm{Red}}+b_{\mathrm{Red}}) \, \pm  \sqrt{ (1-a_{\mathrm{Red}}-b_{\mathrm{Red}})^2 + 4 |d_{\mathrm{Red}}|^2 }  }{ 2} \, \Big\}.
\end{aligned}
\end{equation}
\end{small}\ignorespaces
Consequently, complete positivity requires $\uplambda^{(C_{\mathrm{Red}})} \geq 0$ for all four eigenvalues. This is equivalent to
\begin{equation}
\begin{aligned}
0 &\le a_{\mathrm{Red}}\le 1, \\
0 &\le b_{\mathrm{Red}}\le 1, \\
|c_{\mathrm{Red}}|^2
&\le a_{\mathrm{Red}}(1-b_{\mathrm{Red}}), \\
|d_{\mathrm{Red}}|^2
&\le (1-a_{\mathrm{Red}})b_{\mathrm{Red}}.
\end{aligned}
\label{qubitCPconditions}
\end{equation}
These conditions make clear the distinction between positivity of the instantaneous generator and complete positivity of the finite-time map. The Redfield generator generically possesses a negative instantaneous Kossakowski eigenvalue, yet the integrated evolution need not immediately violate any of the inequalities in Eq.~(\ref{qubitCPconditions}). Conversely, nothing in the Redfield resummation guarantees that these inequalities remain satisfied at arbitrary times. Complete positivity of the Redfield map is therefore a genuinely finite-time question that must be checked from the evolution of $a_{\mathrm{Red}}(t),b_{\mathrm{Red}}(t),c_{\mathrm{Red}}(t)$ and $d_{\mathrm{Red}}(t)$.

The Redfield map evolves via Eq.~(\ref{RedfieldEq_intermsof_Phi2_QUBIT}), and therefore the evolution of its parameters follows directly from Eq.~(\ref{Phi2_supermatrix}). For the population sector one finds
\begin{equation}
\frac{\exd}{\exd t}
\left[
\begin{matrix}
a_{\mathrm{Red}} \\ b_{\mathrm{Red}}
\end{matrix}
\right]
=
-2\lambda^2 \mathrm{Re}[ f_{+} + f_{-} ]  \left[ \begin{matrix} a_{\mathrm{Red}} \\ b_{\mathrm{Red}} \end{matrix} \right]
+
2\lambda^2 \mathrm{Re}[f_{-}] \left[ \begin{matrix}
 1 \\ 1
\end{matrix} \right] \, , \label{ab_redfield_eom}
\end{equation}
while the coherence sector evolves according to
\begin{equation}
\frac{\exd}{\exd t}
\left[
\begin{matrix} c_{\mathrm{Red}} \\ d_{\mathrm{Red}} \end{matrix}
\right] = \lambda^2 \left[ \begin{smallmatrix} - f_{-} - f_{+}^{\ast} & e^{2 i \omega t} [ f_{-} + f_{+}^{\ast} ] \\
e^{-2 i \omega t} [ f_{-}^{\ast} + f_{+} ]
& - f_{-}^{\ast} - f_{+} 
\end{smallmatrix}
\right]
\left[
\begin{matrix}
c_{\mathrm{Red}} \\ d_{\mathrm{Red}}
\end{matrix}
\right] \, , \label{cd_redfield_eom}
\end{equation}
subject to the initial conditions from Eq.~(\ref{abcd_IC}). 

\subsection*{Exact properties of the Redfield map solution}

Some progress can be made in understanding the solutions to Eqs.~(\ref{ab_redfield_eom}) and (\ref{cd_redfield_eom}). The equations of motion can be used to show that the combinations $a_{\mathrm{Red}}-b_{\mathrm{Red}}$ and $|c_{\mathrm{Red}}|^2  - |d_{\mathrm{Red}}|^2$ evolve in a simple way:
\begin{small}
\begin{equation}
\frac{ \frac{\mathrm{d}}{\mathrm{d} t} (a_{\mathrm{Red}}-b_{\mathrm{Red}} ) }{ (a_{\mathrm{Red}}-b_{\mathrm{Red}} ) } = \frac{ \frac{\mathrm{d}}{\mathrm{d} t} (|c_{\mathrm{Red}}|^2-|d_{\mathrm{Red}}|^2 ) }{ (|c_{\mathrm{Red}}|^2-|d_{\mathrm{Red}}|^2 ) } = - 2\lambda^2 \mathrm{Re}[f_{-} + f_{+}],
\end{equation}
\end{small}\ignorespaces
with $(a_{\mathrm{Red}}-b_{\mathrm{Red}})(0) = ( |c_{\mathrm{Red}}|^2 - |d_{\mathrm{Red}}|^2 )(0) = 1$ as the initial conditions. The solution is
\begin{equation}
a_{\mathrm{Red}}(t) - b_{\mathrm{Red}}(t)  = |c_{\mathrm{Red}}(t)|^2 - |d_{\mathrm{Red}}(t)|^2 = e^{- \lambda^2 \Gamma(t)}, \label{aminusb_exact}
\end{equation}
where we have defined
\begin{equation}
\Gamma(t) := 2 \int_0^t \exd s \; \mathrm{Re}[ f_{-}(s) + f_{+}(s) ] \ . \label{Gamma_fpm_def}
\end{equation}
This gives the exact solutions for the populations:
\begin{eqnarray}
a_{\mathrm{Red}}(t) & = & b_{\mathrm{Red}}(t) + e^{- \lambda^2 \Gamma(t)} \label{a_exact}, \\
b_{\mathrm{Red}}(t) & = &  2 \lambda^2 \, e^{- \lambda^2 \Gamma(t)}  \int_0^t \exd s \; e^{\lambda^2 \Gamma(s)} \mathrm{Re}[ f_{-}(s) ] . \qquad \label{b_exact}
\end{eqnarray}
The functions $c_{\mathrm{Red}}(t)$ and $d_{\mathrm{Red}}(t)$ controlling the coherences are more difficult to determine, since their equations remain coupled and time-dependent. Thus, while the combination $|c_{\mathrm{Red}}|^2-|d_{\mathrm{Red}}|^2$ is known exactly, no analogous closed-form expressions for $c_{\mathrm{Red}}(t)$ and $d_{\mathrm{Red}}(t)$ exist for general $f_\pm(t)$.

The exact relation in Eq.~(\ref{aminusb_exact}) also considerably simplifies the conditions for complete positivity. Since $e^{-\lambda^2\Gamma(t)}>0$ one necessarily has $a_{\mathrm{Red}}>b_{\mathrm{Red}}$. Moreover,
\begin{equation}
a_{\mathrm{Red}}(1-b_{\mathrm{Red}})-|c_{\mathrm{Red}}|^2
=
(1-a_{\mathrm{Red}})b_{\mathrm{Red}}-|d_{\mathrm{Red}}|^2 \ ,
\end{equation}
where Eq.~(\ref{aminusb_exact}) has been used in the equality. The two coherence conditions in Eq.~(\ref{qubitCPconditions}) are therefore equivalent, and the complete-positivity conditions reduce to
\begin{equation}
0\leq b_{\mathrm{Red}}\leq a_{\mathrm{Red}}\leq1
\quad\&\quad
 (1-a_{\mathrm{Red}})b_{\mathrm{Red}} \geq |d_{\mathrm{Red}}|^2 \ .
\label{qubitCPconditions2}
\end{equation}
Equivalently, the final inequality may be written as $|c_{\mathrm{Red}}|^2\leq a_{\mathrm{Red}}(1-b_{\mathrm{Red}})$. 

\subsection{Magnus map for a qubit}

It is illuminating to contrast the above Redfield evolution with the Magnus map $\Phi_{\mathrm{Mag}}(t) = e^{ \lambda^2 \Phi_2(t) }$. This approximates the Redfield evolution when the quantity $\epsilon_{\mathrm{Mag}}$ from Eq.~(\ref{epMag_def}) is small.  Our goal in this subsection is to explicitly check that the above map is completely positive. 

The superoperator appearing in the exponential here is the {\it integral} of the matrix $\dot{\Phi}_2(t)$ from Eq.~(\ref{Phi2_supermatrix}). Our first step is to define the definite integrals over functions appearing in Eq.~(\ref{Phi2_supermatrix}):
\begin{eqnarray}
\Lambda_{\pm}(t) & := & \int_0^{t} \exd s\; 2\mathrm{Re}[f_{\pm}(s)], \label{Lambdapm_Def} \\
\Upsilon(t) & = & \int_0^t \exd s\; e^{-2 i \omega s}[ f^{\ast}_{-}(s) + f_{+}(s) ], \\
\Theta(t) & = & - \int_0^{t} \exd s\; \mathrm{Im}[f_{-}(s) + f^{\ast}_{+}(s)].
\end{eqnarray}
We may express
\begin{equation}
\Gamma(t) = \Lambda_{+}(t) + \Lambda_{-}(t)
\end{equation}
using the definition from Eq.~(\ref{Gamma_fpm_def}). Hence 
\begin{equation}
\Phi_2 = \left[
\begin{matrix}
- \Lambda_{+} & \Lambda_{-} & 0 & 0 \\ 
\Lambda_{+} & - \Lambda_{-} & 0 & 0 \\
0 & 0 & - \frac{\Gamma}{2} + i \Theta  & \Upsilon^{\ast} \\
0 & 0 & \Upsilon & - \frac{\Gamma}{2} - i \Theta 
\end{matrix}
\right]  \ . \label{Phi2_Lambdapm}
\end{equation}
Exponentiating this matrix gives a structure analogous to Eq.~(\ref{PhiRed_BLOCKqubit}):
\begin{equation}
\Phi_{\mathrm{Mag}} = \left[
\begin{matrix}
a_{\mathrm{Mag}} & b_{\mathrm{Mag}} & 0 & 0 \\ 
1-a_{\mathrm{Mag}} & 1-b_{\mathrm{Mag}} & 0 & 0 \\
0 & 0 & c_{\mathrm{Mag}} & d_{\mathrm{Mag}}^{\ast} \\
0 & 0 & d_{\mathrm{Mag}} & c_{\mathrm{Mag}}^{\ast}
\end{matrix}
\right],  \label{PhiMag_abcdMag}
\end{equation}
with
\begin{eqnarray}
a_{\mathrm{Mag}}(t) & = & 1 - \frac{\Lambda_{+}(t)}{\Gamma(t)} \bigg[ 1 - e^{ - \lambda^2 \Gamma(t)} \bigg]. \label{aMag_exact} \\
b_{\mathrm{Mag}}(t) & = & \frac{\Lambda_{-}(t)}{\Gamma(t)} \bigg[ 1 - e^{ - \lambda^2 \Gamma(t)} \bigg], \label{bMag_exact}
\end{eqnarray}
and 
\begin{eqnarray}
c_{\mathrm{Mag}}(t) & = & e^{- \tfrac{\lambda^2 \Gamma(t) }{2} }  \bigg[ \cosh\big[ \lambda^2 \sqrt{ |\Upsilon(t)|^2 - \Theta^2(t) } \,  \big] \label{cMag_exact} \\
&& \hspace{18mm} +  i \tfrac{ \Theta(t) \, \sinh\big[ \lambda^2 \sqrt{ |\Upsilon(t)|^2 - \Theta^2(t)  }  \big]}{\sqrt{ |\Upsilon(t)|^2 - \Theta^2(t)  }  } \bigg], \quad \notag \\
d_{\mathrm{Mag}}(t) & = & e^{- \tfrac{\lambda^2 \Gamma(t)}{2} }  \tfrac{ \Upsilon(t) \, \sinh\big[ \lambda^2 \sqrt{ |\Upsilon(t)|^2 - \Theta^2(t)  }  \big]}{\sqrt{ |\Upsilon(t)|^2 - \Theta^2(t) } } .\label{dMag_exact}
\end{eqnarray}
Note that $|\Upsilon|^2 - \Theta^2$ can have either sign, and so the latter two solutions above can take a trigonometric form instead of hyperbolic form.

We see that the Redfield solutions $a(t)$, $b(t)$ from Eqs.~(\ref{a_exact}) and (\ref{b_exact}) reduce to $a_{\mathrm{Mag}}(t)$, $ b_{\mathrm{Mag}}(t)$ when
\begin{equation}
\frac{\Lambda_{\pm}(t)}{\Gamma(t)} \simeq \mathrm{constant} \ .
\end{equation}
Importantly notice that reduction does not require that $\Gamma(t)$ needs to have Markovian-like behaviour. In particular, $\Gamma(t)$ may exhibit non-monotonic or oscillatory time dependence, including intervals where $\dot{\Gamma}(t) < 0$, even as the ratios $\Lambda_{\pm}(t) / \Gamma(t)$ remain constant. This means that agreement between the Redfield and Magnus population dynamics is compatible with non-Markovian evolution, which we do indeed see in later examples.

The question remains: how can we explicitly see that the above solution is completely positive? Since $\Phi_{\mathrm{Mag}}$ in Eq.~(\ref{PhiMag_abcdMag}) is clearly of the same structure as $\Phi_{\mathrm{Red}}$ in Eq.~(\ref{PhiRed_BLOCKqubit}), this means that its corresponding Choi matrix has eigenvalues of the same form as those in Eq.~(\ref{CRed_eigs}). Furthermore, one may check that the Magnus map exactly satisfies
\begin{equation}
a_{\mathrm{Mag}} - b_{\mathrm{Mag}} = |c_{\mathrm{Mag}}|^2 - |d_{\mathrm{Mag}}|^2 \ .
\end{equation}
This means that for the same reasons as for the Redfield map, complete positivity is satisfied when ({\it cf.}~Eq.~(\ref{qubitCPconditions2}))
\begin{equation}
0\leq b_{\mathrm{Mag}}\leq a_{\mathrm{Mag}}\leq1
\quad\&\quad
(1-a_{\mathrm{Mag}})b_{\mathrm{Mag}} \geq |d_{\mathrm{Mag}}|^2 \ . \label{qubitCPconditions_Mag}
\end{equation}
With the Magnus map written in the form of Eqs.~(\ref{aMag_exact})-(\ref{dMag_exact}) it is not yet possible to prove that the complete positivity conditions (\ref{qubitCPconditions_Mag}) are satisfied. As emphasized earlier, proving their satisfaction requires the additional fundamental assumption (\ref{KernelQUBIT}) on the positivity of the environment correlator:
\begin{eqnarray} 
\mathcal E(t,s) = \sum_\kappa
u_{\kappa}(t)
u_{\kappa}^{*}(s) \ .
\end{eqnarray}
It is convenient to define the integrated quantity
\begin{equation}
U_{\pm}^{\kappa}(t) := \int_0^t \exd s \; u_{\kappa}(s) e^{\pm i \omega s},
\end{equation}
in terms of which
\begin{equation} \label{LambdaUpsilon_Ukappa}
\Lambda_{\pm} = \sum_{\kappa} | U_{\pm}^{\kappa} |^2 \qquad  \mathrm{and} \qquad 
\Upsilon = \sum_{\kappa} U_{-}^{\kappa} U_{+}^{\kappa\ast} \ . 
\end{equation}
Notice that this implies (whether the sum over $\kappa$ is discrete or continuous) 
\begin{equation}
\Lambda_{\pm}(t) \geq 0 \qquad \implies \qquad \Gamma(t) \geq 0 \ .
\end{equation}
This does not require $\Gamma(t)$ to be monotonic: although $\Gamma(t)$ is non-negative at every time, its derivative $\dot{\Gamma}(t)$ may have either sign, allowing for the non-Markovian behaviour discussed above.

The above implies $1 - e^{-\lambda^2 \Gamma(t)}  \geq 0$ and therefore $1 - a_{\mathrm{Mag}} \geq 0 $ and $b_{\mathrm{Mag}} \geq 0$, which means the leftmost inequalities in Eq.~(\ref{qubitCPconditions_Mag}) are satisfied. To check the remaining inequality, we compute
\begin{eqnarray}
&& (1-a_{\mathrm{Mag}})b_{\mathrm{Mag}} -  |d_{\mathrm{Mag}}|^2 \\
&& \ \ = \tfrac{\Lambda_{+}\Lambda_{-} ( 1 - e^{ - \lambda^2 \Gamma} )^2 }{\Gamma^2}  - e^{-\lambda^2\Gamma} |\Upsilon|^2  \Big| \tfrac{ \sinh[ \lambda^2 \sqrt{ |\Upsilon|^2 - \Theta^2 } ] }{ \sqrt{ |\Upsilon|^2 - \Theta^2 } } \Big|^2 \notag \\
&& \ \ = \lambda^4 |\Upsilon|^2 e^{- \lambda^2 \Gamma } \bigg[ \tfrac{\Lambda_{+}\Lambda_{-}}{|\Upsilon|^2} \tfrac{ \sinh^2( \frac{\lambda^2 \Gamma}{2} ) }{ ( \frac{\lambda^2\Gamma}{2} )^2 } - \Big| \tfrac{ \sinh[ \lambda^2 \sqrt{ |\Upsilon|^2 - \Theta^2 } ] }{ \lambda^2 \sqrt{ |\Upsilon|^2 - \Theta^2 } } \Big|^2 \bigg]. \notag
\end{eqnarray}
So far this is an equality, where we must keep the modulus-square of the second term involving the square root to cover both cases $|\Upsilon|^2 - \Theta^2 \gtrless 0$. 

Crucially, the decomposition from Eq.~(\ref{LambdaUpsilon_Ukappa}) tells us that
\begin{small}
\begin{equation} 
\bigg| \sum_{\kappa}  U_{-}^{\kappa} U_{+}^{\kappa\ast} \bigg|^2 \leq \sum_{\kappa,\overline{\kappa}} | U_{-}^{\kappa} |^2 | U_{+}^{\overline{\kappa}} |^2 \ \implies \frac{ \Lambda_{+} \Lambda_{-} }{|\Upsilon|^2} \geq 1,
\end{equation}
\end{small}\ignorespaces
using the Cauchy-Schwarz inequality. This implies that 
\begin{eqnarray}
&& (1-a_{\mathrm{Mag}})b_{\mathrm{Mag}} -  |d_{\mathrm{Mag}}|^2 \\
&& \quad \geq \lambda^4 |\Upsilon|^2 e^{- \lambda^2 \Gamma } \bigg[ \tfrac{ \sinh^2( \frac{\lambda^2 \Gamma}{2} ) }{ ( \frac{\lambda^2\Gamma}{2} )^2 } - \Big| \tfrac{ \sinh[ \lambda^2 \sqrt{ |\Upsilon|^2 - \Theta^2 } ] }{ \lambda^2 \sqrt{ |\Upsilon|^2 - \Theta^2 } } \Big|^2 \bigg] \ . \notag
\end{eqnarray}
The quantity in the square brackets turns out to always be non-negative\footnote{To see this one must consider both $|\Upsilon|^2-\Theta^2\gtrless0$. When $|\Upsilon|^2-\Theta^2>0$, one notes that $\sqrt{|\Upsilon|^2-\Theta^2}\leq|\Upsilon|\leq\sqrt{\Lambda_+\Lambda_-}$. Moreover, $(\Lambda_+-\Lambda_-)^2\geq0$ implies $\Gamma^2\geq4\Lambda_+\Lambda_-$, and hence $\sqrt{|\Upsilon|^2-\Theta^2}\leq\Gamma/2$. The desired result then follows from the fact that $\sinh(x)/x$ is monotonically increasing for $x\geq0$. When instead $\Theta^2-|\Upsilon|^2>0$, the second term analytically continues to $\sin^2(y)/y^2$. In this case the result follows immediately from $\sin^2(y)/y^2\leq1\leq\sinh^2(x)/x^2$ for real $y$ and $x\geq0$. This covers both cases (the former which relies on Cauchy-Schwarz).}, and we conclude at last that $(1-a_{\mathrm{Mag}})b_{\mathrm{Mag}} -  |d_{\mathrm{Mag}}|^2\geq 0$, showing explicilty that the Magnus map is completely positive.

Finally, we ask when qubit Redfield evolution is approximately the same as Magnus evolution, the latter of which we've shown is completely positive in two ways. The approximation is controlled by the hierarchy $\epsilon_{\mathrm{Mag}} = \lambda^2 | \Omega_{4} | / | \Omega_2 | \ll 1$, where the definition (\ref{epMag_def}) involves the computation of
\begin{eqnarray}
\Omega_{4}(t) &=& \frac{1}{2} \int_{0}^{t} \exd s\, [ \dot{\Phi}_2(s), \Phi_2(s) ]  \\
&=& \left[
\begin{array}{cccc}
\mathcal{X}(t) & \mathcal{X}(t) & 0 & 0 \\ 
- \mathcal{X}(t) & - \mathcal{X}(t) & 0 & 0 \\
0 & 0 & i \mathcal{Y}(t) & \mathcal{Z}^{\ast}(t) \\
0 & 0 & \mathcal{Z}(t) & - i \mathcal{Y}(t)
\end{array}
\right], \label{omega4_matXYZ}
\end{eqnarray}
with the definitions
\begin{eqnarray}
\mathcal{X}(t) & = & \frac{1}{2} \int_0^t \exd s \; \big( \Lambda_{+}(s) \dot{\Lambda}_{-}(s) - \Lambda_{-}(s) \dot{\Lambda}_{+}(s) \big)\ , \qquad \\
\mathcal{Y}(t) & = & \int_0^t \exd s \; \mathrm{Im}\big[ \Upsilon(s) \dot{\Upsilon}^{\ast}(s) \big] \ , \\
\mathcal{Z}(t) & = &  i  \int_{0}^t \exd s \; \bigg[ \Theta(s) \dot{\Upsilon}(s) - \Upsilon(s) \dot{\Theta}(s) \bigg] \ ,
\end{eqnarray}
where we've used $\Phi_2$ from Eq.~(\ref{Phi2_Lambdapm}). Note that $\mathcal{X}$ and $\mathcal{Y}$ are real, while $\mathcal{Z}$ is complex. Taking the
norm in Eq.~(\ref{epMag_def}) to be the Frobenius norm, which is
particularly convenient for the qubit case, we obtain
\begin{equation}
\epsilon_{\mathrm{Mag}}(t)
=
\lambda^2
\sqrt{
\frac{
2\mathcal{X}^2
+\mathcal{Y}^2
+|\mathcal{Z}|^2
}{
\frac{5\Lambda_{+}^2 + 2\Lambda_{+}\Lambda_{-}  + 5\Lambda_{-}^2}{4} +|\Upsilon|^2 +\Theta^2  } } \ .
\label{epsilonMag_qubit}
\end{equation}
Asking for this quantity to be small depends on the physics of the particular problem at hand, in particular the nature of the environmental degrees of freedom and their interactions with the system. To this end we now explore two examples.

\section{A qubit coupled to a qubit}
\label{sec:qubitqubit}

We will first illustrate the preceding discussion using arguably the simplest microscopic model: a qubit coupled to another qubit. This example has several useful features. The exact reduced dynamics can be computed analytically, the environmental correlation function has an elementary positive-kernel decomposition, and the various perturbative resummations can be compared directly.

We consider the Hamiltonian with terms
\begin{equation}
H_{\mathrm{S}}
=
\frac{\omega}{2}\sigma_{3},
\qquad
H_{\mathrm{E}}
=
\frac{\Omega}{2}\sigma_{3},
\qquad
H_{\mathrm{int}}
=
\lambda\,\sigma_{1}\otimes\sigma_{1}.
\label{qubitqubit_ex}
\end{equation}
and assume $\Omega \ge \omega$ for convenience. In the notation of Eq.~(\ref{generalqubit_Hint}), the interaction-picture interaction
$H_{\mathrm{int}}^{I}(t)=\sigma_1(t)\otimes E(t)$ has
\begin{equation}
E(t)
=
\sigma_{+}e^{+i\Omega t}
+
\sigma_{-}e^{-i\Omega t}.
\end{equation}
We take the initial system-environment state to be
\begin{equation}
\rho(0)
=
\varrho_{\mathrm{Si}}
\otimes
\ket{\downarrow}\bra{\downarrow},
\end{equation}
so that the environment qubit begins in its ground state.

\subsection*{Exact vs. approximate reduced dynamics}

Since the full system consists of only two qubits, its dynamics can be solved exactly. For a simple comparison of the different approximation schemes, we specialize to an initially-maximally mixed system,
\begin{equation}
\varrho_{\mathrm{Si}}
=
\frac{\mathbb{I}}{2}\ . \label{maxmixed_in}
\end{equation}
The exact reduced state remains diagonal
\begin{equation}
\varrho_{\mathrm{S}}(t)
= \Phi(t) \bigg[ \frac{\mathbb{I}}{2} \bigg] =  
\left[
\begin{matrix}
p(t) & 0 \\
0 & 1-p(t)
\end{matrix}
\right]\ , \label{exactqq_ans}
\end{equation}
with
\begin{equation} \label{pexact_qubitqubit}
p(t)
=
\frac{1}{2}
-\tfrac{\lambda^2 \sin^2( \sqrt{ \lambda^2 + \delta^2_{-}/4 }\; t ) }{2 [ \lambda^2 + \delta_{-}^2/4  ] } + \tfrac{\lambda^2 \sin^2( \sqrt{ \lambda^2 + \delta_{+}^2/4 }\; t ) }{2 [ \lambda^2 + \delta_{+}^2/4] },
\end{equation}
where we have defined
\begin{equation} \label{deltapm_def}
\delta_{\pm}:=\Omega \pm \omega  \ .
\end{equation}
On resonance, $\Omega=\omega$, this simplifies to
\begin{equation}
p(t) \big|_{\Omega = \omega}
=
\frac{1}{2}\cos^2(\lambda t)
+
\frac{\lambda^2 \sin^2\big(\sqrt{\lambda^2+\omega^2}\; t\big) }{2(\lambda^2+\omega^2)} \ .
\label{pexact_resonance}
\end{equation}
The exact reduced quantum map $\Phi(t)$ can also be obtained analytically\footnote{The exact map $\Phi(t)$ from Eq.~(\ref{exactqq_ans}) has the same form as $\Phi_{\mathrm{Red}}$ and $\Phi_{\mathrm{Mag}}$, with its coefficients saturating both nontrivial complete-positivity inequalities with
$|c|^2=a(1-b)$ and $|d|^2=(1-a)b$.} and is CPTP (as it must be).

Our next goal is to approximate the population function $p(t)$ appearing in
Eq.~(\ref{exactqq_ans}) using the different schemes introduced above:
ordinary perturbation theory and the resolvent, Redfield, and Magnus
resummations. For the maximally-mixed initial state, each of these
approximations preserves the diagonal form of the reduced density matrix,
but gives a different prediction for $p(t)$. 

The environmental correlation function entering the second-order reduced dynamics is particularly simple:
\begin{equation}
\mathcal{E}(t,s)
=
\bra{\downarrow}E(t)E(s)\ket{\downarrow}
=
e^{-i\Omega(t-s)}.
\label{Equbitqubit}
\end{equation}
It has the rank-one positive-kernel decomposition
\begin{equation}
\mathcal{E}(t,s)
=
u(t)u^{\ast}(s)
\qquad \mathrm{with} \ 
u(t)=e^{-i\Omega t},
\end{equation}
so that the positivity condition discussed above is manifest. For this correlation function, the functions $f_{\pm}(t)$ from Eq.~(\ref{fpm_def}) evaluate to
\begin{equation}
f_{\pm}(t)
=
\frac{1-e^{-i(\Omega\mp \omega)t}}
{i(\Omega \mp \omega)} = \frac{1-e^{-i\delta_{\mp} t}}
{i\delta_{\mp}} \ .
\label{fpm_qubitqubit}
\end{equation}
One then integrates Eq.~(\ref{Lambdapm_Def}) to find
\begin{equation}
\Lambda_{\pm}(t)=
\frac{4\sin^2(\frac{\delta_{\mp}}{2} t)}{\delta_{\mp}^2} \ .
\label{Lambda_qubitqubit}
\end{equation}
For the maximally-mixed initial state, ordinary second-order perturbation theory obtained via $\Phi_{2}$ in Eq.~(\ref{Phi2_Lambdapm}) gives the population evolution
\begin{eqnarray}
p_{\mathrm{pert}}(t)
& = &
\frac{1}{2}
-
\frac{\lambda^2}{2}
\left[ \Lambda_+(t) - \Lambda_-(t)
\right] \\
& =  & \frac{1}{2}
-\frac{ 2 \lambda^2 \sin^2( \tfrac{\delta_{-}}{2} t ) }{ \delta_{-}^2 } + \frac{ 2 \lambda^2 \sin^2( \tfrac{\delta_{+}}{2} t ) }{ \delta_{+}^2 }. \qquad \label{ppert_qubitqubit}
\end{eqnarray}
Indeed this is the perturbative limit of Eq.~(\ref{pexact_qubitqubit}).  We see that perturbation theory is valid in the limit
\begin{equation}
\frac{\lambda}{\Omega - \omega} \ll 1,\label{PT_validqubit}
\end{equation}
so that the first term in Eq.~(\ref{ppert_qubitqubit}) is bounded.
This breakdown becomes particularly transparent at exact resonance $\Omega=\omega$ where the condition above cannot be satisfied. In this limit,
\begin{equation}
p_{\mathrm{pert}}(t) \big|_{\Omega = \omega}
=
\frac{1}{2}
-\frac{\lambda^2t^2}{2}
+
\frac{\lambda^2\sin^2(\omega t)}{2\omega^2}.
\label{ppert_resonance}
\end{equation}
which clearly demonstrates secular growth (as can also be seen by expanding the first term in Eq.~(\ref{pexact_resonance})). The secular term eventually drives the perturbative density matrix outside the space of positive states at times $t\sim\mathcal{O}(\lambda^{-1})$.

\begin{figure}[t]
\begin{center}
\includegraphics[width=0.45\textwidth]{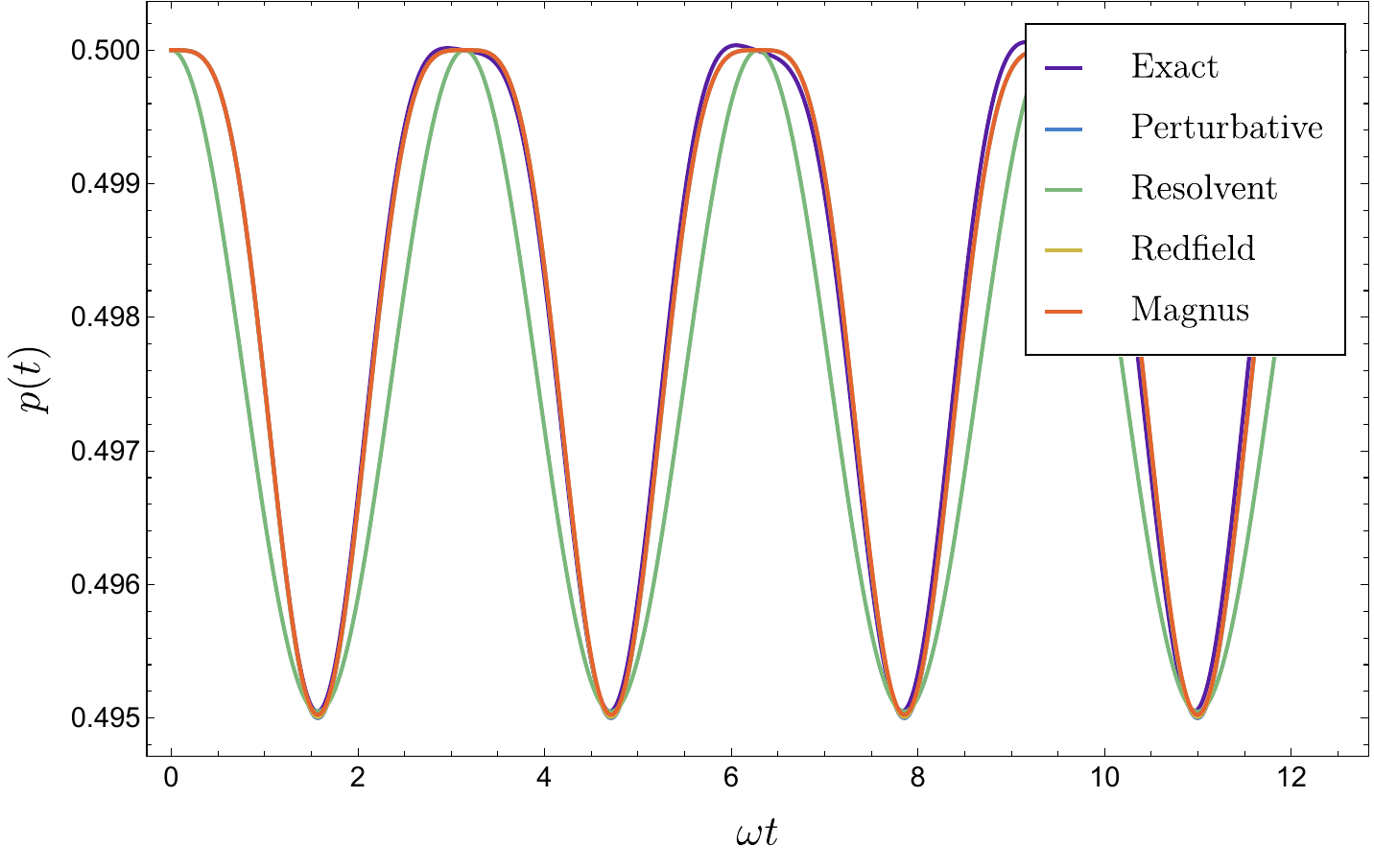}
\hspace{3mm}
\includegraphics[width=0.45\textwidth]{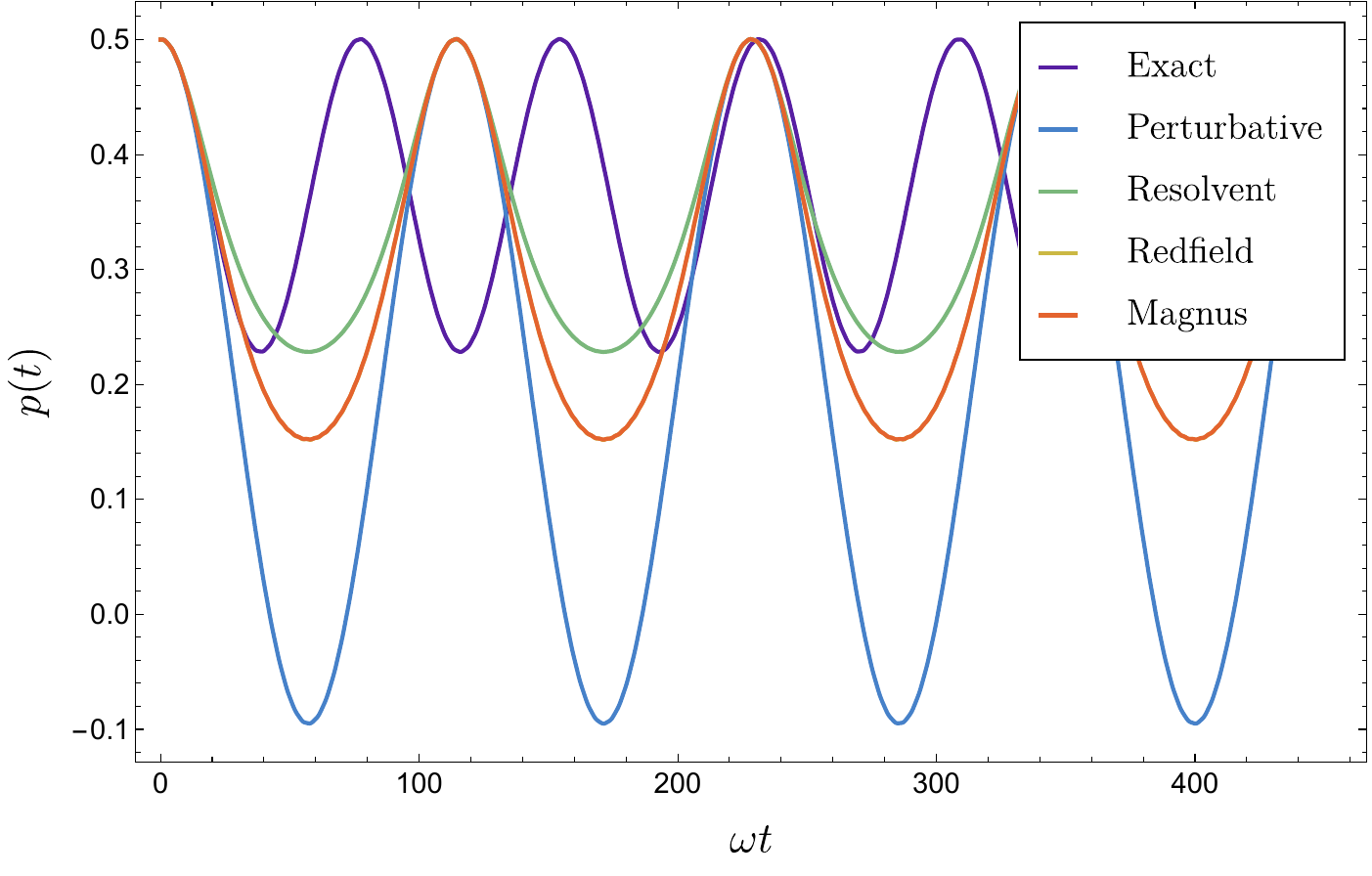}
\caption{\small
Evolution of the excited-state population for an initially maximally-mixed system qubit (\ref{maxmixed_in}). We compare the exact evolution with second-order perturbation theory and the resolvent, Redfield, and Magnus resummations. The top panel shows a detuned example with $\Omega/\omega = 3$ and $\lambda/\omega = 0.1$. The bottom panel shows a near resonance case $\Omega\simeq\omega$ with $\Omega/\omega = 1.055$ and $\lambda/\omega = 0.03$, in which even perturbation theory does a poor job at approximating the exact dynamics. The Redfield and Magnus resummations are nearly indistinguishable at this scale; they are compared more directly in Figure \ref{Fig:qubitqubit_CP}. Notice that the resolvent resummation remains CP but does a poor job at approximating the exact dynamics.}
\label{Fig:qubitqubit_comparison}
\end{center}
\end{figure}

The resolvent resummation $\Phi_{\mathrm{res}}
= (
\mathcal{I}-\lambda^2\Phi_2
)^{-1}$ from Eq.~(\ref{ResolventDefinition}) instead predicts a population
\begin{equation}
p_{\mathrm{res}}(t) = \frac{\frac{1}{2 } + \lambda^2 \Lambda_{-}(t)}{1 + \lambda^2 [  \Lambda_{+}(t) + \Lambda_{-}(t) ] }.
\end{equation}
On resonance this becomes
\begin{equation}
p_{\mathrm{res}}(t) \big|_{\Omega = \omega}
= \frac{\frac{1}{2} + \frac{\lambda^2 \sin^2(\omega t)}{\omega^2} }{ 1 + \lambda^2 t^2 + \frac{\lambda^2 \sin^2(\omega t)}{\omega^2} },
\label{pres_qubitqubit}
\end{equation}
which algebraically resums the earlier secular growth. Finally, the Redfield map $\Phi_{\mathrm{Red}}$ in Eq.~(\ref{PhiRed_BLOCKqubit}) predicts that the population obeys 
\begin{equation}
\frac{\exd p_{\mathrm{Red}}}{\exd t}
=
\lambda^2\dot{\Lambda}_-(t)
-
\lambda^2\dot{\Gamma}(t)p_{\mathrm{Red}}(t) \quad \mathrm{with} \ 
p_{\mathrm{Red}}(0)=\frac{1}{2},
\label{pRedODE_qubitqubit}
\end{equation}
which has the solution
\begin{equation}
p_{\mathrm{Red}}(t)
= \frac{1}{2} - \frac{\lambda^2 e^{- \lambda^2 \Gamma(t)}}{2}  \int_0^t \exd s \; e^{\lambda^2\Gamma(s)}
[ \dot{\Lambda}_+(s) - 
\dot{\Lambda}_-(s) ] \ .
\label{pRed_qubitqubit}
\end{equation}
where we recall that $\Gamma = \Lambda_{+} + \Lambda_{-}$. The Magnus resummation via $\Phi_{\mathrm{Mag}}$ from Eq.~(\ref{PhiMag_abcdMag}) tells us instead
\begin{equation}
p_{\mathrm{Mag}}(t)
=
\frac{1}{2}
-
\frac{\Lambda_+(t) - \Lambda_-(t)}
{2\Gamma(t)}
\left[
1-e^{-\lambda^2\Gamma(t)}
\right] \ .
\label{pMag_qubitqubit}
\end{equation}
One can check that all the resummations reproduce the perturbative result to leading order in the coupling. We compare the exact dynamics directly with ordinary perturbation theory and the resolvent, Redfield, and Magnus resummations in Figure \ref{Fig:qubitqubit_comparison}.

\subsection*{Approximating Redfield as Magnus}

The preceding comparison concerns the evolution of one particular initial state. We now ask the stronger question of when the Redfield map itself approximates the Magnus map.
We first compute the coefficients appearing in $\Phi_2$ from Eq.~(\ref{Phi2_Lambdapm})
\begin{eqnarray}
\Upsilon(t)
&=&
\frac{
4e^{-i\omega t}
\sin(\frac{\delta_+t}{2})
\sin(\frac{\delta_-t}{2})
}{
\delta_+\delta_-
},
\\
\Theta(t)
&=&
\frac{t}{\delta_+}
-\frac{\sin(\delta_+t)}{\delta_+^2}
-\frac{t}{\delta_-}
+\frac{\sin(\delta_-t)}{\delta_-^2},
\end{eqnarray}
where the $\Lambda_{\pm}$ are given in Eq.~(\ref{Lambda_qubitqubit}). The integrals appearing in Eq.~(\ref{omega4_matXYZ}) can then be computed exactly; for example,
\begin{equation}
\mathcal{X}(t) = -\frac{2 (\Omega  \sin (t \omega )-\omega  \sin (t \Omega ))^2}{\omega  \Omega  \left(\omega ^2-\Omega ^2\right)^2}.
\end{equation}
$\mathcal{X}(t)$ stays bounded at late times, but $\mathcal{Y}(t)$ and $\mathcal{Z}(t)$ grow at late times:
\begin{eqnarray}
\mathcal{Y}(t) & \simeq & \frac{4 \omega }{\left(\Omega ^2-\omega ^2\right)^2} \, t + \ldots \\
| \mathcal{Z}(t) | & \simeq & \tfrac{2 \omega  \sqrt{(\cos (\omega t )+2 \cos (\Omega t ))^2+9 \sin ^2(\omega t  )}}{\left(\Omega ^2-\omega ^2\right)^2} \, t  + \ldots  \qquad
\end{eqnarray}
Finally, we compute the parameter $\epsilon_{\mathrm{Mag}}$ from Eq.~(\ref{epsilonMag_qubit}). At late times the dominant behaviour is
\begin{eqnarray}
\epsilon_{\mathrm{Mag}}(t) & \simeq & \lambda^2 \tfrac{\sqrt{(\cos (\omega t  )+2 \cos (\Omega t  ))^2+9 \sin ^2(\omega t )+4}}{\Omega ^2-\omega ^2} \\
& \lesssim & \sqrt{ \frac{35}{2} }\; \frac{\lambda^2}{\Omega ^2-\omega ^2}. \label{ep_mag_qubqub}
\end{eqnarray}
It is this quantity that should be small in order for the Redfield map to approximate the Magnus map. Near resonance, $\Omega-\omega\ll\omega$, Eq.~(\ref{ep_mag_qubqub})
implies
\begin{equation}
\epsilon_{\mathrm{Mag}}
\lesssim
\sqrt{\frac{35}{8}}\,
\frac{\lambda^2}{\omega(\Omega-\omega)}.
\end{equation}
Comparing this with the condition
$\lambda/(\Omega-\omega)\ll1$ from Eq.~(\ref{PT_validqubit}) required for fixed-order perturbation
theory, we find an intermediate regime
\begin{equation}
\frac{\lambda^2}{\omega}
\ll
\Omega-\omega
\lesssim
\lambda\ , \label{PTbrok_RedMag} 
\end{equation}
in which fixed-order perturbation theory is no longer uniformly
controlled, while the Redfield and Magnus resummations may nevertheless
remain close.

This separation of scales is already illustrated in
Fig.~\ref{Fig:qubitqubit_comparison}. In the near-resonant example of the bottom panel,
perturbation theory already provides a poor approximation
to the exact population dynamics, while the Redfield and Magnus
resummations remain nearly indistinguishable.  This is precisely the intermediate regime identified above: the perturbative expansion is no longer uniformly controlled, even though the two resummations remain close.

\begin{figure}[t]
\begin{center}
\includegraphics[width=0.45\textwidth]{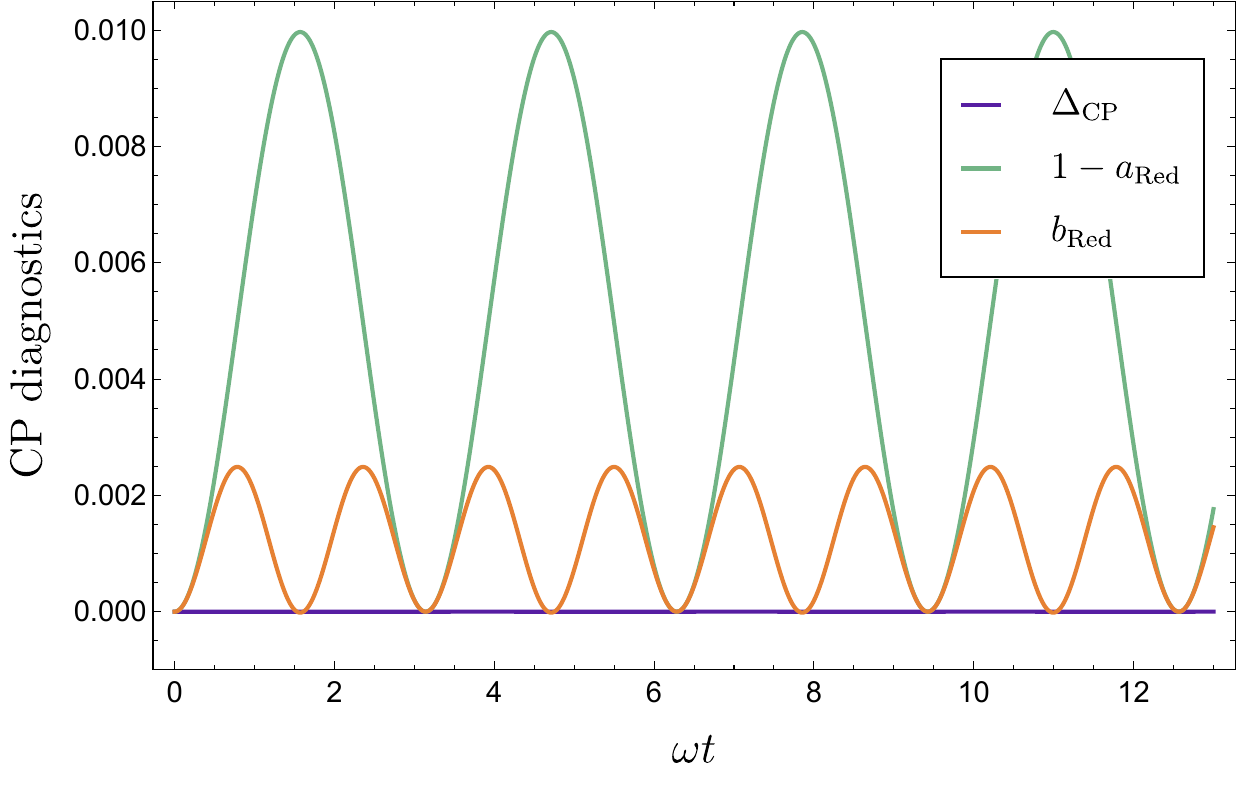}
\hspace{3mm}
\includegraphics[width=0.45\textwidth]{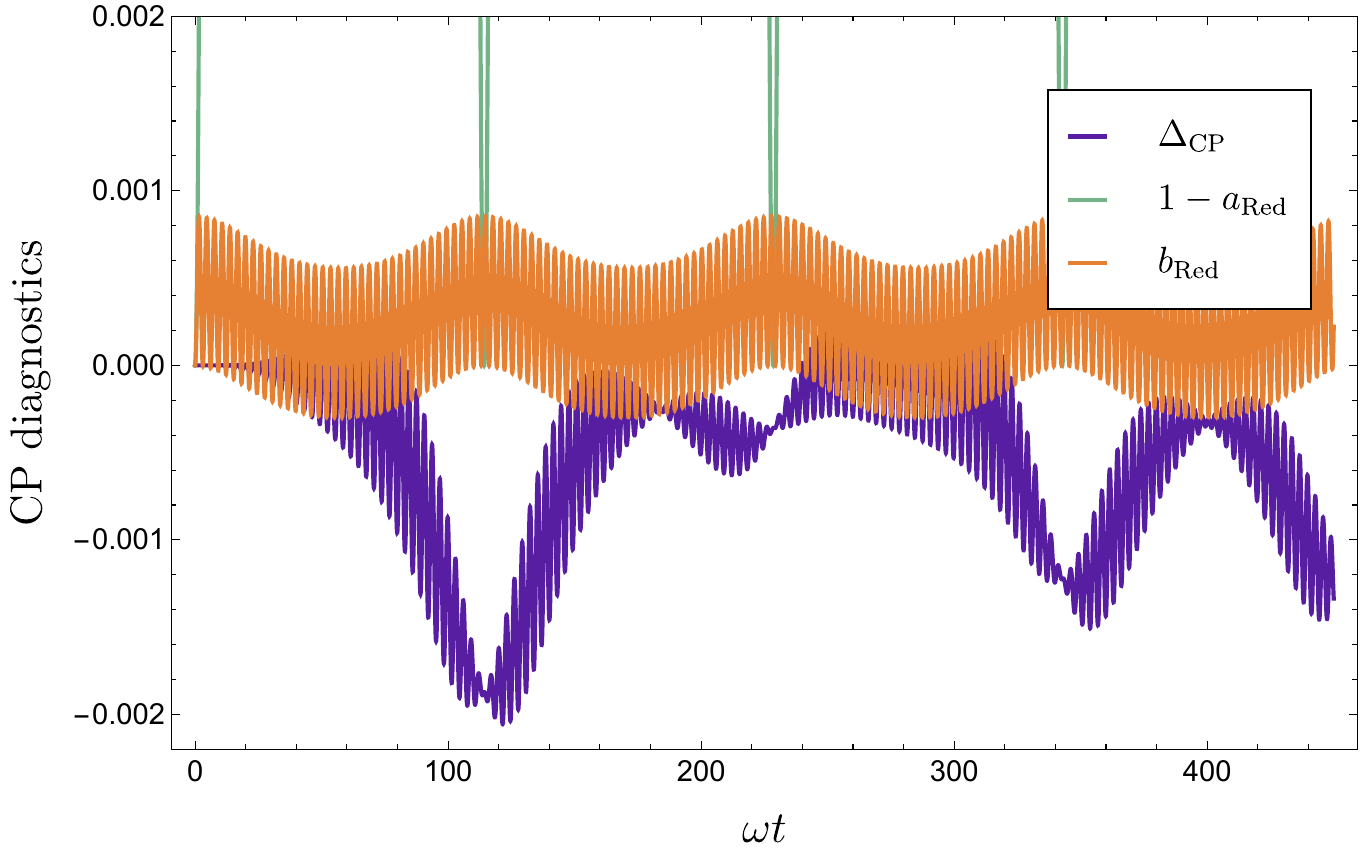}
\caption{\small Complete-positivity diagnostics for the Redfield map, evaluated for
the same parameter choices as in
Fig.~\ref{Fig:qubitqubit_comparison}. The top panel shows the
detuned case, $\Omega/\omega=3$ and $\lambda/\omega=0.1$, while the
bottom panel shows the near-resonant case, $\Omega/\omega=1.055$ and
$\lambda/\omega=0.03$.  The Redfield map remains CPTP only while all three quantities are non-negative. Small violations develop in the near-resonant example, even though the Redfield and Magnus population dynamics are nearly indistinguishable.}
\label{Fig:qubitqubit_CP}
\end{center}
\end{figure}

Agreement for a particular initial state does not, however, guarantee
that the full Redfield map is CPTP. We therefore show in
Fig.~\ref{Fig:qubitqubit_CP} the three independent complete-positivity diagnostics (see Eq.~(\ref{qubitCPconditions2})
\begin{equation}
1-a_{\mathrm{Red}} \; ,\quad
b_{\mathrm{Red}}\; ,\quad
\Delta_{\mathrm{CP}}
=
[1-a_{\mathrm{Red}}]b_{\mathrm{Red}}
-
|d_{\mathrm{Red}}|^2 .
\end{equation}
For the detuned example, all three remain non-negative over the range
shown.  Near resonance, by contrast, the Redfield map develops small violations of complete positivity, despite its population dynamics
remaining extremely close to those generated by the Magnus map. The
Magnus map itself remains CPTP by construction.

In summary, $\epsilon_{\mathrm{Mag}}$
controls the proximity of the Redfield and Magnus resummations, with the latter providing a CPTP map.  Importantly, the condition for this proximity can be parametrically weaker than the condition required for the validity fixed-order perturbation theory as in Eq.~(\ref{PTbrok_RedMag}). There is consequently a regime in
which the perturbative expansion has already broken down, while
Redfield remains close to a CPTP resummation and its violations of
complete positivity are correspondingly small.

\section{A qubit coupled to damping oscillators}
\label{sec:qubit_damp}

We conclude by considering a more non-trivial example, where one may be in perturbative control whilst not necessarily having Redfield$\simeq$Magnus. In this example, we couple our qubit to a collection of harmonic osillators such that
\begin{equation} \label{osc_bath}
H_{\mathrm{S}} = \frac{\omega}{2}\sigma_{3} \quad \mathrm{and} \quad H_{\mathrm{E}}
=
\sum_{k}
\varepsilon_{k} \Big(
a_{k}^{\dagger}a_{k}  + \tfrac{1}{2} \Big),
\end{equation}
where $a_k$ and $a_k^{\dagger}$ are ladder operators satisfying $[a_k,a_q^{\dagger}] = \delta_{kq} \mathbb{I}_{\mathrm{E}}$. We choose the well-known interaction\footnote{Choosing $H_{\mathrm{int}} = \lambda\, \sum_{k} g_{k} ( \sigma_{+} \otimes a_{k} + \sigma_{-} \otimes a_{k}^{\dagger} )$ instead would conserve the total excitation number, letting us solve the model exactly. However, in this setting the Redfield and Magnus maps are identical, since $[\Phi_2, \dot{\Phi}_2] =0$ exactly, so we don't explore this choice further.} of the spin-boson model:
\begin{equation}
H_{\mathrm{int}} = \lambda\, \sum_{k} g_{k} \, \sigma_1 \otimes (a_k + a_k^{\dagger}) .
\end{equation}
%
Here we keep $g_{k}$ as dimensionless, so that $\lambda$ retains dimensions of energy like in the previous example.

\begin{figure}[h]
\begin{center}
\includegraphics[width=80mm]{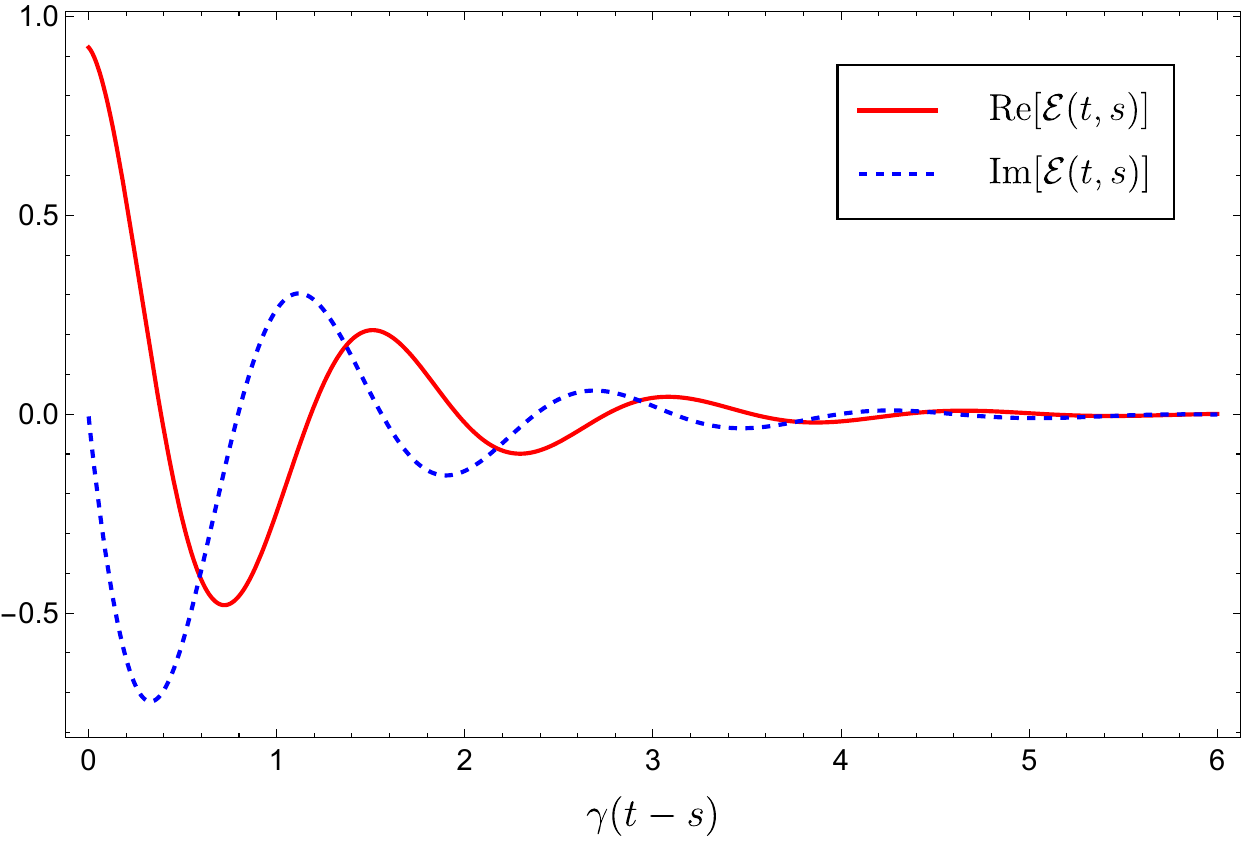}
\caption{\small
Plot of the correlator $\mathcal{E}(t,s)$ from Eq.~(\ref{spectralEts}) for the choice $\Omega / \gamma = 4$. Notice that the fall-off of the imaginary part is much slower than the real part, consistent with Eq.~(\ref{falloff}).}
\label{Fig:CorrPlot}
\end{center}
\end{figure}

We choose the system to start in an initial pure state and the environment to start in the vacuum:
\begin{equation}
\rho(0) = \varrho_{\mathrm{Si}} \otimes  | 0 \rangle \langle 0 | ,
\end{equation}
i.e.\ $a_{k}|0\rangle = 0$ for each $k$. The environmental correlation function is therefore
\begin{equation}
\mathcal E(t,s)
= \sum_{k} 
g_k^2 \,  
e^{-i\varepsilon_k(t-s)}.
\end{equation}
In the continuum limit\footnote{For example, take uniformly spaced frequencies
$\varepsilon_k= k \Delta\varepsilon$ with $k=0,1,2,\ldots$ and couplings
$g_k^2= \frac{\Delta\varepsilon}{\pi} \frac{  \gamma }{(k\Delta \varepsilon - \Omega)^2+\gamma^2}$. As $\Delta\varepsilon\to 0$ the sum over $k$ becomes the Riemann integral in Eq.~(\ref{spectralEts}).}, we choose the couplings $g_k$ and energies $\varepsilon_{k}$ so that the above correlator has a Lorentzian spectral density,
\begin{equation}
\mathcal E(t,s)
\simeq 
\int_{0}^{\infty}
\frac{d\varepsilon}{\pi}
\;
\frac{\gamma}
{(\varepsilon - \Omega)^2+\gamma^2}\, 
e^{-i\varepsilon(t-s)} \;  \label{spectralEts}
\end{equation}
for fixed $\gamma > 0$. For $t - s>0$ this evaluates to
\begin{eqnarray}
\mathcal E(t,s) & \simeq & \frac{e^{(\gamma -i \Omega ) (t-s)}+e^{-(\gamma +i \Omega ) (t-s)}}{2} \\
&& + \frac{i}{2\pi}\Big[ \, e^{(\gamma -i \Omega ) (t-s)} \text{Ei}\big(-(\gamma -i \Omega ) (t-s)\big)\notag \\
&& \qquad -e^{-(\gamma +i \Omega ) (t-s)} \text{Ei}\big((\gamma +i \Omega ) (t-s)\big) \, \Big] ,\notag
\end{eqnarray}
where $\mathrm{Ei}(z) = -  \int_{-z}^{\infty} \exd r\, e^{-r}/r$ is the exponential integral function defined via the Cauchy principal value for real arguments and by analytic continuation elsewhere in the complex plane. We plot the correlator in Figure \ref{Fig:CorrPlot}. Its late-time behaviour is 
\begin{equation} \label{falloff}
\mathcal{E}(t,s) \simeq e^{ - (\gamma +i \Omega ) (t-s)}  -\frac{i \gamma }{\pi(\gamma ^2+\Omega ^2) (t-s)} + \ldots
\end{equation}
for $\gamma (t-s) \gg 1$. In the $\gamma \to 0^{+}$ limit, the correlator takes the especially simple form
\begin{equation}
\mathcal{E}(t,s) \big|_{\Omega =0} = e^{- i \Omega (t-s)},
\end{equation}
since the spectral representation approaches a Dirac delta function near the frequency $\Omega$. In this limit, we thus have the same environment correlator as in the qubit case, Eq.~(\ref{Equbitqubit}). Hence this example contains all the same the non-Markovian behaviour reflected by the earlier qubit example, but gives us additional control over a damping parameter.

Defining
\begin{equation}
u_{\varepsilon}(t) = \sqrt{ \frac{\gamma}{\pi[ ( \varepsilon - \Omega)^2 + \gamma^2]} }\; e^{ - i \varepsilon t}
\end{equation}
lets us decompose the spectral representation (\ref{spectralEts}) to make positivity manifest:
\begin{equation}
\mathcal{E}(t,s) = \int_{-\infty}^{\infty} \exd \varepsilon \; u_{\varepsilon}(t) u^{\ast}_{\varepsilon}(s),
\end{equation}
where the label $\varepsilon$ is continuous. 

Now we can ask when the Redfield evolution is approximately equivalent to Magnus. We first compute the $f_{\pm}(t)$, defined in Eq.~(\ref{fpm_def}):
\begin{eqnarray}
&& f_{\pm}(t) = \frac{1 -  e^{-(\gamma +i (\Omega \mp \omega ))t}}{2 [ \gamma +i (\Omega \mp \omega )]} + \frac{e^{(\gamma -i (\Omega \mp \omega ))t} \mp 1}{2[\gamma -i (\Omega \mp \omega )]} \qquad \\
&& \quad + \frac{i e^{\pm i \omega t }}{2 \pi } \Big[ \, \tfrac{e^{-(\gamma +i \Omega )t} \text{Ei}((\gamma +i \Omega ) t)}{\gamma +i (\Omega \mp \omega )}+\tfrac{e^{(\gamma -i \Omega )t} \text{Ei}(-(\gamma -i \Omega ) t)}{\gamma -i (\Omega \mp \omega )} \, \Big] \notag \\
&& \quad -\frac{i}{2 \pi } \Big[ \, \tfrac{\log (\frac{\gamma -i \Omega }{\omega }) \pm \frac{i \pi }{2}}{\gamma -i (\Omega \mp \omega )}+\tfrac{\log(\frac{\gamma +i \Omega }{\omega }) \mp \frac{i \pi }{2}}{\gamma + i (\Omega \mp \omega )}+\tfrac{2 \gamma  \text{Ei}(\pm i \omega  t)}{\gamma ^2+(\Omega \mp \omega )^2} \, \Big]. \notag
\end{eqnarray}
%

To make progress analytically, we must also compute the functions $\Lambda_{\pm}$, $\Upsilon$, $\Theta$ which appear in Eq.~(\ref{Phi2_Lambdapm}) for $\Phi_2$. Using the spectral representation (\ref{spectralEts}) of the correlator, the time integrals can be performed:
\begin{eqnarray}
\Lambda_{\pm}(t) & = & \frac{4\gamma}{\pi} \int_0^\infty \exd \varepsilon\;  \frac{ \sin^2\big( \frac{\varepsilon \mp \omega}{2} t \big) }{[(\varepsilon - \Omega)^2 + \gamma^2](\varepsilon \mp \omega)^2}, \\
\Upsilon(t) & = & \frac{2\gamma e^{- i \omega t}}{\pi} \int_0^\infty \exd \varepsilon \; \frac{\cos(\omega t) - \cos(\varepsilon t)}{[(\varepsilon - \Omega)^2 + \gamma^2](\varepsilon^2 - \omega^2)}, \qquad \ \  \\
\Theta(t) & = & \frac{\gamma}{\pi}
\int_0^\infty
\frac{\exd\varepsilon}{(\varepsilon-\Omega)^2+\gamma^2}
 \\
&& \ \times \bigg[
\frac{2\omega t}{\omega^2-\varepsilon^2} + \frac{\sin[(\varepsilon-\omega)t]}{(\omega-\varepsilon)^2}
- \frac{\sin[(\omega+\varepsilon)t]}{(\omega+\varepsilon)^2} \bigg]. \notag
\end{eqnarray}
In the late-time limit $\gamma t \gg 1$,
\begin{eqnarray}
&& \Lambda_{+}(t) \simeq  \tfrac{2\gamma}{\gamma^2 + (\Omega - \omega)^ 2 } \,  t + \ldots \\
&& \Lambda_{-}(t) \simeq  \tfrac{2 \gamma}{\pi \omega (\gamma ^2+(\Omega +\omega )^2)} + \nu + \ldots \\
&& \Upsilon(t) \simeq \gamma  e^{-i \omega t } \Big[ \tfrac{\sin ( \omega t)}{\omega(\gamma ^2+(\Omega -\omega )^2)} + \mu \cos(\omega t) \Big] + \ldots \qquad \\
&& \Theta(t) \simeq - \gamma \omega \mu t + \ldots 
\end{eqnarray}
where we define the coefficients
\begin{eqnarray}
\mu & := & \tfrac{2}{\pi} \tfrac{\gamma  \Omega  \log\big(\frac{\gamma ^2+\Omega ^2}{\omega ^2}\big)-(\gamma ^2+\omega ^2-\Omega ^2) [\tan ^{-1}(\frac{\Omega }{\gamma })+\frac{\pi }{2}]}{\gamma  \left(\gamma ^2+(\Omega -\omega )^2\right) \left(\gamma ^2+(\Omega +\omega )^2\right)}, \\
\nu  & := & \tfrac{2}{\pi} \tfrac{\gamma ( \Omega + \omega )  \log\big(\frac{\gamma ^2+\Omega ^2}{\omega ^2}\big)-(\gamma ^2 - (\Omega + \omega)^2) [\tan ^{-1}(\frac{\Omega }{\gamma })+\frac{\pi }{2}]}{\left(\gamma ^2+(\Omega +\omega )^2\right)^2}. \qquad
\end{eqnarray}
We can use the above asymptotics to find the late-time limits of the integrals appearing in Eq.~(\ref{omega4_matXYZ}):
\begin{eqnarray}
\mathcal{X}(t) & \simeq & -\tfrac{\gamma}
{\gamma^2+(\Omega-\omega)^2}
\Big[
\tfrac{2\gamma}
{\pi\omega(\gamma^2+(\Omega+\omega)^2)}
+\nu
\Big]t+\ldots \\
\mathcal{Y}(t) & \simeq & \tfrac{\gamma^2\omega}{2}
\left[
\tfrac{1}
{\omega^2(\gamma^2+(\Omega-\omega)^2)^2}
+\mu^2
\right]t+\ldots \\
\mathcal{Z}(t) & \simeq & \gamma^2\mu 
\Big[
\tfrac{\cos(\omega t)}
{\gamma^2+(\Omega-\omega)^2}
-\omega \mu\sin(\omega t) \Big] t e^{-i\omega t}
+\ldots \qquad \ 
\end{eqnarray}
All three of these functions grow secularly at late times. In particular the envelope of the oscillations of $\mathcal{Z}(t)$ is bounded by 
\begin{equation}
|\mathcal{Z}(t)|^2 \lesssim \gamma^4\mu^2 
\Big[
\tfrac{1}
{(\gamma^2+(\Omega-\omega)^2)^2}
+\omega^2 \mu^2 \Big] t^2 \ .
\end{equation}
Inserting the above expressions into Eq.~(\ref{epsilonMag_qubit}) one gets a cumbersome expression bounding $\epsilon_{\mathrm{Mag}}$ at late times. It is this expression $\epsilon_{\mathrm{Mag}} \ll 1$ which must be made small. We will not write the full general expression, but only in a few limits.

First consider the limit of short correlation time, where we find
\begin{equation} \label{eMag_mark}
\epsilon_{\mathrm{Mag}} \lesssim \sqrt{\tfrac{1}{20}+\tfrac{8}{5 \pi ^2}}\; \frac{\lambda^2}{\omega \gamma}  \ll 1  \qquad \mathrm{for \ } \gamma \gg \Omega, \omega.
\end{equation}
This is the regime normally associated with Markovianity, although here the oscillations ruin any semigroup structure.

In the opposite, long correlation-time limit, we find
\begin{equation} \label{eMag_nonmark}
\epsilon_{\mathrm{Mag}} \lesssim  \frac{\sqrt{5} \lambda^2}{|\Omega^2 - \omega^2|}  \ll 1  \qquad \mathrm{for \ } \gamma \ll \Omega, \omega
\end{equation}
Notice that this limit encapsulates the scaling seen for the environment qubit in the previous example: in the limit $\gamma \to 0^{+}$, one cannot be too close to resonance.

\begin{figure}[t]
    \centering
   \includegraphics[width=0.95\columnwidth]{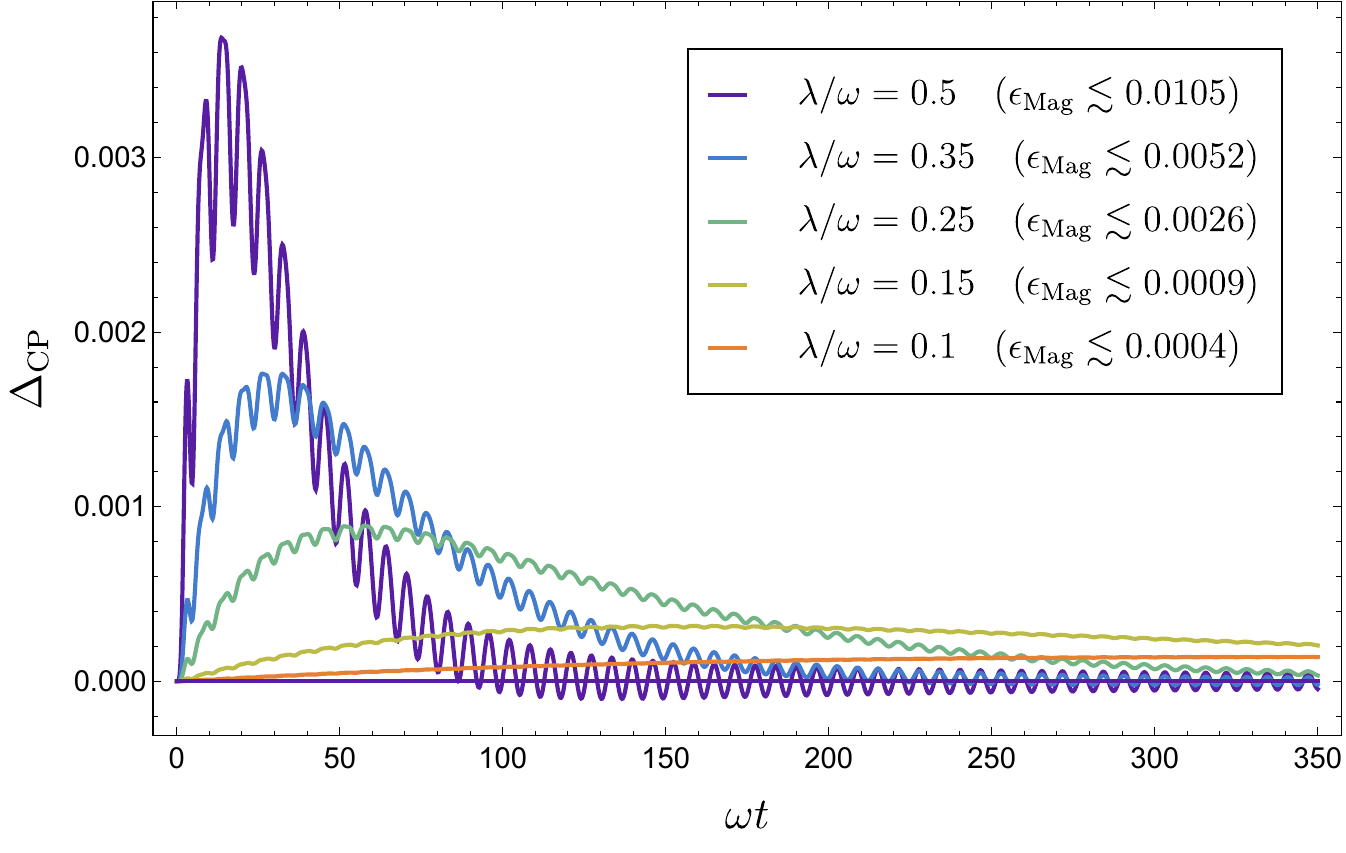}

    \vspace{2mm}

    \includegraphics[width=0.95\columnwidth]{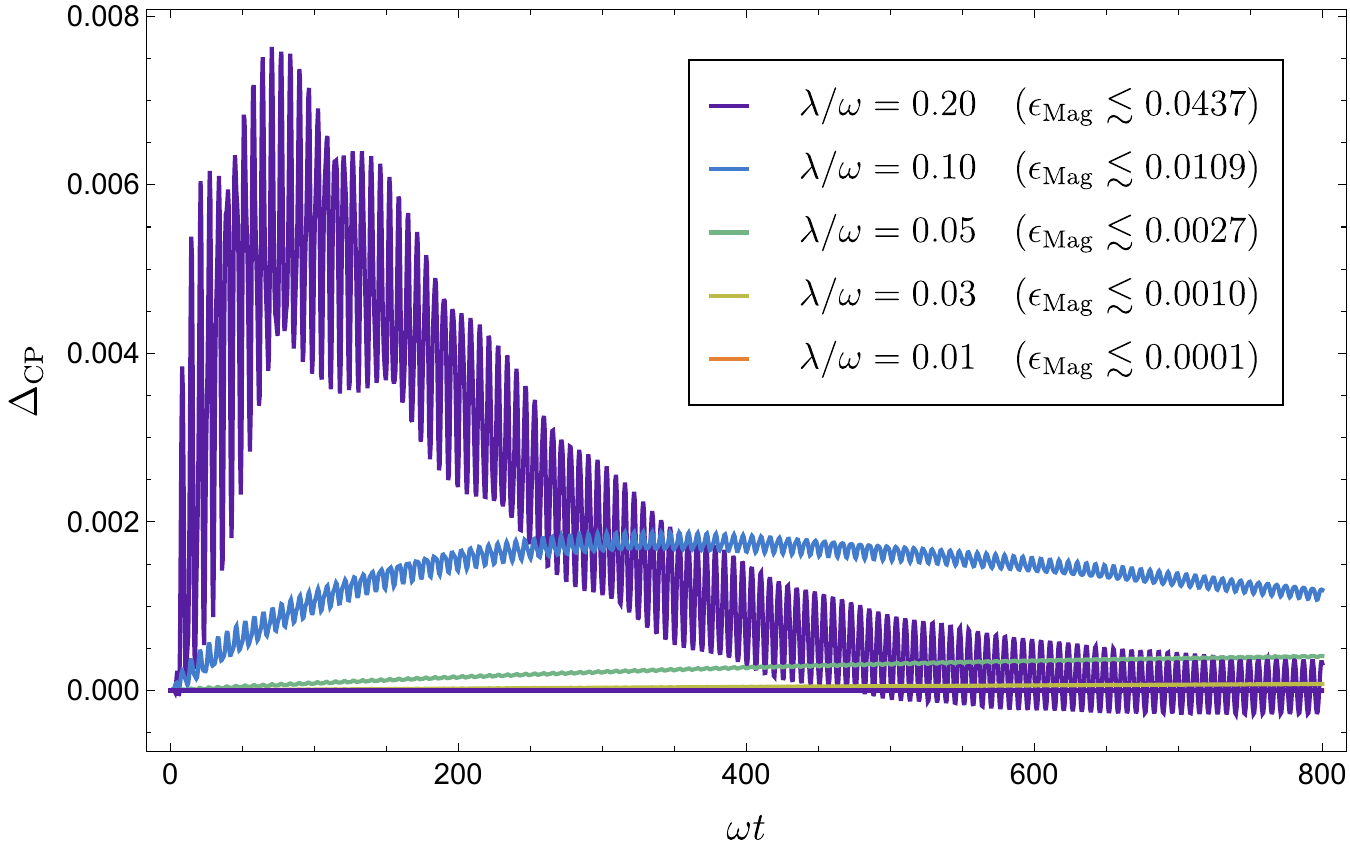}

    \vspace{2mm}

    \includegraphics[width=0.95\columnwidth]{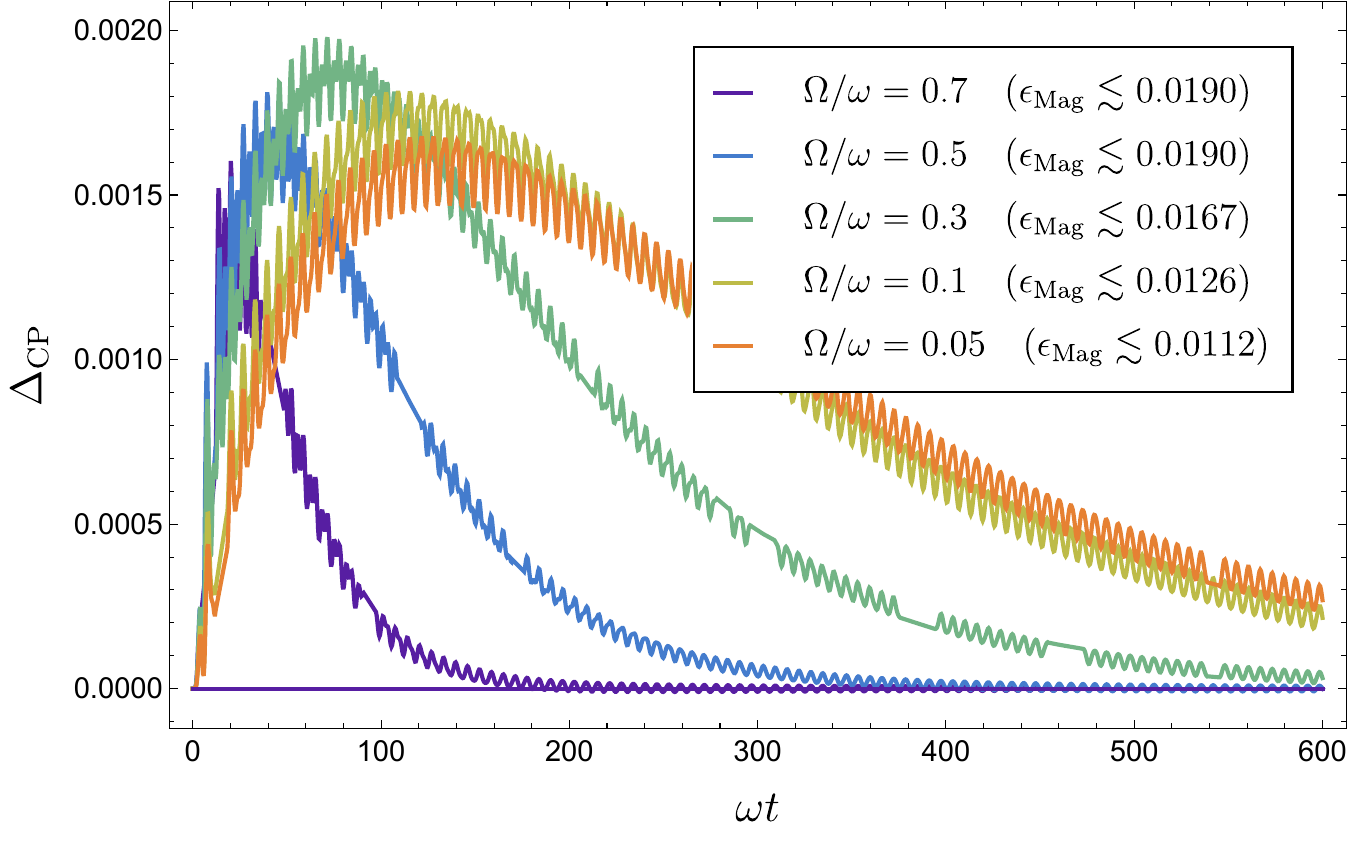}

    \caption{
    Complete-positivity diagnostic $\Delta_{\mathrm{CP}}$ as a function of $\omega t$ for the Redfield map in three parameter regimes. The top panel varies $\lambda/\omega$, with $\Omega/\omega=0$ and $\gamma/\omega=10$ fixed. The middle panel varies $\lambda/\omega$, with $\Omega=0$ and $\gamma/\omega=0.1$ fixed. The bottom panel varies $\Omega/\omega$, with $\lambda/\omega=0.1$ and $\gamma/\omega=0.25$ fixed. In each case, increasing the late-time value $\epsilon_{\mathrm{Mag}}$ is accompanied by increasing violations of complete positivity at late times.
    }
    \label{fig:Ex2_CP_scans}
\end{figure}

We now illustrate this criterion numerically. In
Fig.~\ref{fig:Ex2_CP_scans}, we plot the complete-positivity diagnostic
\begin{equation}
\Delta_{\mathrm{CP}}(t)
=
[1-a_{\mathrm{Red}}(t)]b_{\mathrm{Red}}(t)
-
|d_{\mathrm{Red}}(t)|^2
\end{equation}
for several choices of parameters.
We focus on examples for which the violations of complete positivity occur at late times, where the asymptotic estimate of $\epsilon_{\mathrm{Mag}}$ is applicable. In each of the three parameter scans, larger values of the late-time  $\epsilon_{\mathrm{Mag}}$ are accompanied by larger violations of complete positivity. Conversely, as $\epsilon_{\mathrm{Mag}}$ becomes small, the Redfield map approaches the manifestly CPTP Magnus map and the violations are correspondingly suppressed.

We have deliberately restricted Fig.~\ref{fig:Ex2_CP_scans} to examples where the relevant violations occur in the late-time regime. More generally, the Redfield map can also violate complete positivity during initial transient behavior, before the asymptotic expressions leading to {\it eg.}~Eqs.~(\ref{eMag_mark}) and (\ref{eMag_nonmark}) apply. Describing such cases would
require following the full time-dependent quantity
$\epsilon_{\mathrm{Mag}}(t)$ rather than its late-time limit. We do not
pursue a more detailed transient analysis here.

In closing, we emphasize that $\epsilon_{\mathrm{Mag}}$ controls only the first time-ordering correction distinguishing the Redfield and
Magnus resummations, rather than providing a bound on the complete
Magnus series. Nevertheless, it provides a useful diagnostic of when
the Redfield map approaches the manifestly CPTP Magnus map, as
illustrated by Fig.~\ref{fig:Ex2_CP_scans}.

\section{Conclusions}

The moral of this paper is simple: the road from perturbation theory to GKLS is paved with good intentions (and with resummations), but GKLS is not the only sensible destination. Weak-coupling perturbation theory fixes the reduced dynamics only order by order, and very often only at early times. Promoting this perturbative information to a master equation or other resummed evolution necessarily introduces higher-order terms. It is in this sense that resummation is an art: different resummations may agree perturbatively while behaving very differently once formally higher-order terms become important.

We have studied the Redfield equation with time-dependent coefficients starting from this perspective. For the positive environmental correlation functions considered here, the second-order perturbative map admits manifestly CPTP resummations, including the resolvent and Magnus maps. In particular, the difference between Redfield and Magnus evolution is controlled by superoperator time-ordering. We introduced the diagnostic $\epsilon_{\mathrm{Mag}}$ indicating when Redfield remains close to the CPTP Magnus map. Our examples show that Redfield can remain well behaved despite having a non-GKLS instantaneous generator, including in regimes where ordinary fixed-order perturbation theory has already broken down.

The broader lesson is to distinguish the perturbative information supplied by the microscopic theory from the resummation used to extend it. Complete positivity provides an important guide, but neither instantaneous GKLS form nor semigroup structure is required for sensible open-system evolution. Different resummations have different regimes of validity, and the most important question is therefore not which one has the most desirable formal properties, but rather when the higher-order terms it implicitly introduces remain under control. We hope that viewing Redfield evolution through this lens makes it a little less worrying, and perhaps even a little easier to love.

\section*{Acknowledgements}

We thank Xiantao Li, Corey Ostrove, Sarah Shandera and Erickson Tjoa for useful discussions.

OpenAI GPT-5.6 Sol, accessed through ChatGPT, was used to search for relevant references, explore connections with the existing literature, check intermediate algebraic calculations and formulas, and assist with the generation and debugging of Mathematica code. All results and calculations presented in this work were verified by the authors, who take full responsibility for the content and mathematical statements of the manuscript.

The Mathematica notebooks used for the calculations and figures in this work are publicly available as ancillary files accompanying the arXiv submission.

\begin{appendix}

\section{Time-ordered exponentials and the Magnus expansion}
\label{App:Magnus}

For completeness, we briefly summarize the time-ordered exponential and Magnus expansion used in the main text. Consider the linear evolution equation
\begin{equation}
\frac{\exd}{\exd t}\Phi(t,t_{\mathrm i})
=
A(t)\Phi(t,t_{\mathrm i}),
\qquad
\Phi(t_{\mathrm i},t_{\mathrm i})=\mathcal I .
\end{equation}
Its formal solution is the time-ordered exponential
\begin{equation}
\Phi(t,t_{\mathrm i})
=
\mathcal T
\exp\left[
\int_{t_{\mathrm i}}^t \exd s\,A(s)
\right],
\end{equation}
defined by the Dyson series
\begin{eqnarray}
\Phi(t,t_{\mathrm i})
&=&
\mathcal I
+
\int_{t_{\mathrm i}}^t \exd s\,A(s) \label{timeordereddef}
\\
&&
+
\int_{t_{\mathrm i}}^t \exd s
\int_{t_{\mathrm i}}^s \exd u\,
A(s)A(u)
+\cdots .
\notag
\end{eqnarray}
The time-ordering superoperator $\mathcal T$ places operators evaluated at later times to the left.

The same evolution may alternatively be written using the Magnus expansion,
\begin{equation}
\Phi(t,t_{\mathrm i})
=
\exp\left[
\Omega_1(t)+\Omega_2(t)+\Omega_3(t)+\cdots
\right],
\end{equation}
where the first three terms are:
\begin{eqnarray}
\Omega_1(t) & = &
\int_{t_{\mathrm i}}^t \exd s\,A(s) \\
\Omega_2(t)
& = &
\frac{1}{2}
\int_{t_{\mathrm i}}^t \exd s
\int_{t_{\mathrm i}}^s \exd u\,
[A(s),A(u)] \\
\Omega_3(t)
& = & \frac{1}{6}
\int_{t_{\mathrm i}}^t \exd s
\int_{t_{\mathrm i}}^s \exd u
\int_{t_{\mathrm i}}^u \exd v\,  \\
&& \times \Big(
\big[A(s),[A(u),A(v)]\big] +
\big[A(v),[A(u),A(s)]\big]
\Big) \notag
\end{eqnarray}
Thus the first Magnus approximation neglects the effects of time ordering,
\begin{equation}
\mathcal T
\exp\left[
\int_{t_{\mathrm i}}^t \exd s\,A(s)
\right]
\simeq
\exp\left[
\int_{t_{\mathrm i}}^t \exd s\,A(s)
\right].
\end{equation}
In particular, if $[A(s),A(u)]=0$ for all $s,u$ then all higher Magnus terms vanish and this relation becomes exact.

\section{Mean-field contributions}
\label{App:MF}

In Eq.~(\ref{MFvanish}) in the main text we made the simplifying assumption that the one-point functions
\begin{equation}
\overline{E}_{\alpha}(t)
:=
\trB
\left[
E_{\alpha}(t)\varrho_{\mathrm{Ei}}
\right]
\end{equation}
of the environment operators vanish. This is not essential to our results. In this appendix we briefly show how the perturbative, Redfield, and Magnus maps are modified when $\overline{E}_{\alpha}(t)\neq0$.

Beginning from the second-order expansion of the full density matrix $\rho(t)$ from Eq.~(\ref{perturbation_theory}), and making use of the interaction $H^{I}_{\mathrm{int}}(t)$ from Eq.~(\ref{Hint_SE}), after a partial trace one finds that the perturbative map for the reduced density matrix takes the form
\begin{equation}
\Phi(t,t_{\mathrm i})
=
\mathcal I
+
\lambda\Phi_1(t,t_{\mathrm i})
+
\lambda^2\Phi_2(t,t_{\mathrm i})
+
\mathcal O(\lambda^3) \; 
\label{PTmap_MF}
\end{equation}
where now there is a contribution linear in the coupling,
\begin{equation}
\Phi_1(t,t_{\mathrm i})[\varrho_{\mathrm{Si}}]
= -i \int_{t_{\mathrm i}}^t
\exd s\,
\left[
H_{\mathrm{MF}}(s),
\varrho_{\mathrm{Si}}
\right]
\label{Phi1_MF}
\end{equation}
where we've defined the mean-field Hamiltonian
\begin{equation}
H_{\mathrm{MF}}(t)
:=
\sum_{\alpha}
\overline{E}_{\alpha}(t)S_{\alpha}(t) \; .
\label{HMF}
\end{equation}
The second-order contribution $\Phi_2(t,t_{\mathrm i})$ takes a form identical to Eq.~(\ref{Phi2_explicit}) where the environment correlator 
\begin{equation}
\mathcal{E}_{\alpha\beta}(t,s)
:=
\trB
\left[
E_{\alpha}(t)E_{\beta}(s)\varrho_{\mathrm{Ei}}
\right]
\label{Efull_MF}
\end{equation}
contains {\it both} the connected fluctuations and the contribution from the environment mean-field. For this reason it is useful to introduce the centered environment operators
\begin{equation}
\delta E_{\alpha}(t)
:=
E_{\alpha}(t)-\overline{E}_{\alpha}(t),
\end{equation}
and define the centered environment correlator
\begin{eqnarray}
\mathcal E^{\mathrm{c}}_{\alpha\beta}(t,s)
&:=&
\trB
\left[
\delta E_{\alpha}(t)
\delta E_{\beta}(s)
\varrho_{\mathrm{Ei}}
\right]
\notag
\\
&=&
E_{\alpha\beta}(t,s)
-
\overline{E}_{\alpha}(t)
\overline{E}_{\beta}(s).
\label{centeredCORR_MF}
\end{eqnarray}
When the one-point functions vanish, notice how $\mathcal E^{\mathrm{c}}$ reduces to the correlator defined in Eq.~(\ref{plaindefEenv_introduction}) of the main text.

Both $\mathcal E_{\alpha\beta}(t,s)$ and its centered counterpart $\mathcal E^{\mathrm c}_{\alpha\beta}(t,s)$ define positive two-argument kernels. For our purposes, however, it is useful to apply the positive mode decomposition to the centered correlator:
\begin{equation}
\mathcal E^{\mathrm c}_{\alpha\beta}(t,s)
=
\sum_{\kappa}
u_{\kappa\alpha}(t)
u_{\kappa\beta}^{*}(s)] \; .
\label{KernelDecomp_MF}
\end{equation}
This decomposition applies equally when the environment one-point functions are nonzero. Explicit examples of nonvanishing mean-fields for which the corresponding centered correlator is decomposed in this way (directly within a Redfield-TDC equation) can be found in recent cosmological open-system calculations \cite{Burgess:2022nwu,Burgess:2026jyc}.

To make the resulting structure transparent, it is useful to separate the second-order contribution into the part generated by the mean-field Hamiltonian and the part generated by the centered environment fluctuations. To this end we define
\begin{equation}
\mathcal K_2(t,t_{\mathrm i})
:=
\Phi_2(t,t_{\mathrm i})
-
\tfrac{1}{2}\Phi_1^2(t,t_{\mathrm i}) \; .
\label{K2_MF}
\end{equation}
The perturbative map (\ref{PTmap_MF}) may equivalently be written as
\begin{equation}
\Phi(t,t_{\mathrm i})
=
\mathcal I
+
\lambda\Phi_1
+
\lambda^2
\Big(
\mathcal K_2+\tfrac{1}{2}\Phi_1^2
\Big)
+
\mathcal O(\lambda^3) \; .
\label{PTmap_cumulant_MF}
\end{equation}
The reason for introducing $\mathcal K_2$ in this way is that the  $\tfrac{1}{2}\Phi_1^2$ term is already generated at second order by exponentiating the first-order contribution. The combination $\mathcal K_2$ has a particularly simple interpretation. Using $\mathcal E_{\alpha\beta}
=
\mathcal E^{\mathrm c}_{\alpha\beta}
+
\overline E_\alpha\overline E_\beta$, one finds that (\ref{K2_MF}) is equivilent to
\begin{equation}
\mathcal K_2[\varrho_{\mathrm{Si}}]
=
\Phi_2^{\mathrm c}[\varrho_{\mathrm{Si}}] -i
\big[
H_{\mathrm{MF}}^{(2)},\, \varrho_{\mathrm{Si}} 
\big] 
\label{K2split_MF}
\end{equation}
where $\Phi_2^{\mathrm c}$ is exactly the second-order map (\ref{Phi2_explicit}) from the main text with the replacement $\mathcal E_{\alpha\beta}\rightarrow \mathcal E^{\mathrm c}_{\alpha\beta}$, while
\begin{equation}
H_{\mathrm{MF}}^{(2)}(t,t_{\mathrm i})
:=
-\frac{i}{2}
\int_{t_{\mathrm i}}^t \exd s
\int_{t_{\mathrm i}}^s \exd u\,
\left[
H_{\mathrm{MF}}(s),
H_{\mathrm{MF}}(u)
\right] \; .
\label{HMF2}
\end{equation}
The positive decomposition (\ref{KernelDecomp_MF}) then allows $\Phi_2^{\mathrm c}$ to be written in a GKLS form analagous to that in the main text in Eq.~(\ref{Phi2GKLSform}). That is to say, we take ({\it cf.}~Eq.(\ref{Vkappa_def}))
\begin{equation}
V_\kappa(t,t_{\mathrm{i}})
=
\sum_\alpha
\int_{t_{\mathrm i}}^{t}
ds\;
u_{\kappa\alpha}^{*}(s)
S_\alpha(s)\; .
\end{equation}
where $u_{\kappa\alpha}$ are the mode functions from Eq.~(\ref{KernelDecomp_MF}), as well as ({\it cf.}~Eqs.~(\ref{HBdefinition}) and (\ref{Bdef_main}))
\begin{equation}
H^{\mathrm{c}}_B(t,t_{\mathrm{i}})
:=
\frac{B^{\mathrm{c}}(t,t_{\mathrm{i}})-B^{\mathrm{c}\dagger}(t,t_{\mathrm{i}})}{2i} 
\end{equation}
with
\begin{equation}
B^{\mathrm{c}}(t,t_{\mathrm{i}})
=
\sum_{\alpha,\beta}
\int_{t_{\mathrm i}}^{t}
ds
\int_{t_{\mathrm i}}^{s}
du\;
\mathcal E^{\mathrm{c}}_{\beta\alpha}(s,u)
S_\beta(s) 
S_\alpha(u) \ . 
\end{equation}
Consequently, we find
\begin{eqnarray}
(\lambda\Phi_1+\lambda^2\mathcal K_2)
[\varrho_{\mathrm{Si}}] &=&
-i
\big[
H_{\mathrm{eff}},
\varrho_{\mathrm{Si}}
\big] \label{PhiGKLS_MF}
\\
&+& \lambda^2
\sum_\kappa
\Big(
V_\kappa\varrho_{\mathrm{Si}}V_\kappa^\dagger
-
\tfrac{1}{2}
\left\{
V_\kappa^\dagger V_\kappa,
\varrho_{\mathrm{Si}}
\right\}
\Big) \notag
\end{eqnarray}
where
\begin{eqnarray}
H_{\mathrm{eff}}(t,t_{\mathrm i})
& = &
\lambda
\int_{t_{\mathrm i}}^t
\exd s\,
H_{\mathrm{MF}}(s) \label{Heff_MF} \\
&& \hspace{10mm} + \lambda^2
\Big(
H_B^{\mathrm c}(t,t_{\mathrm i})
+
H_{\mathrm{MF}}^{(2)}(t,t_{\mathrm i})
\Big) \notag
\end{eqnarray}
which shows that the presence of a non-vanishing environment mean-field adds Hamiltonian contributions while leaving the dissipative structure unchanged. An approximate Kraus representation can likewise be obtained by extending the analysis of \S\ref{sec:PT}, although we do not write it explicitly here.

We next consider the corresponding Redfield evolution. From Eq.~(\ref{PTmap_MF}), the inverse map has the expansion
\begin{equation}
\Phi^{-1}
=
\mathcal I
-
\lambda\Phi_1
+
\lambda^2
\left(
\Phi_1^2-\Phi_2
\right)
+
\mathcal O(\lambda^3),
\end{equation}
which means that
\begin{equation}
\dot{\Phi}\Phi^{-1}
=
\lambda\dot{\Phi}_1
+
\lambda^2
\left(
\dot{\Phi}_2-\dot{\Phi}_1\Phi_1
\right)
+
\mathcal O(\lambda^3).
\end{equation}
A simple calculation using the decomposition of $\Phi_2$ above shows that
\begin{equation}
\dot{\Phi}_2-\dot{\Phi}_1\Phi_1
=
\dot{\Phi}_2^{\mathrm c}.
\label{TCL2centered_MF}
\end{equation}
The Redfield-TDC equation (\ref{ExactTCL}) therefore takes the particularly simple form
\begin{equation}
\frac{\exd\varrho_{\mathrm S}}{\exd t}
=
\left[
\lambda\dot{\Phi}_1(t,t_{\mathrm i})
+
\lambda^2\dot{\Phi}_2^{\mathrm c}(t,t_{\mathrm i})
\right]
[\varrho_{\mathrm S}(t)] \; ,
\label{Redfield_MF}
\end{equation}
with the corresponding map ({\it cf.}~Eq.~(\ref{RedfieldMap}))
\begin{equation}
\Phi_{\mathrm{Red}}(t,t_{\mathrm i})
=
\mathcal T
\exp
\bigg[
\int_{t_{\mathrm i}}^t
\exd s\,
\big(
\lambda\dot{\Phi}_1(s,t_{\mathrm i})
+
\lambda^2\dot{\Phi}_2^{\mathrm c}(s,t_{\mathrm i})
\big)
\bigg] .
\label{RedfieldMap_MF}
\end{equation}
The non- vanishing mean-field enters the Redfield generator through the first-order Hamiltonian contribution $\lambda\dot{\Phi}_1$, while the second-order contribution is determined entirely by the centered environment fluctuations. When $\overline E_\alpha=0$, one has $\Phi_1=0$ and $\Phi_2^{\mathrm c}=\Phi_2$, recovering the Redfield equation and map used in the main text.

The corresponding CPTP Magnus map is obtained by exponentiating $\lambda\Phi_1+\lambda^2\mathcal K_2$
\begin{equation}
\Phi_{\mathrm{Mag}}(t,t_{\mathrm i})
:=  e^{
\lambda\Phi_1(t,t_{\mathrm i})
+
\lambda^2\mathcal K_2(t,t_{\mathrm i}) } \label{MagnusMap_MF} 
\end{equation}
The exponent is of GKLS form by Eq.~(\ref{PhiGKLS_MF}), and so $\Phi_{\mathrm{Mag}}$ is CPTP. Using Eq.~(\ref{K2_MF}) one finds
\begin{eqnarray}
\Phi_{\mathrm{Mag}} & = & e^{
\lambda\Phi_1
+
\lambda^2 (
\Phi_2-\tfrac{1}{2}\Phi_1^2
 ) } \notag \\ 
& \simeq &
\mathcal I
+
\lambda\Phi_1
+
\lambda^2\Phi_2
+
\mathcal O(\lambda^3) \ ,
\end{eqnarray}
so $\Phi_{\mathrm{Mag}}$ agrees with ordinary perturbation theory through second order. For vanishing mean-fields, $\Phi_1=0$ and $\mathcal K_2=\Phi_2$, this all recovers Eq.~(\ref{MagnusMap}) of the main text.

Finally, the relation between $\Phi_{\mathrm{Red}}$ and $\Phi_{\mathrm{Mag}}$ follows directly from the Magnus expansion. Writing the Redfield map (\ref{RedfieldMap_MF}) as
\begin{eqnarray}
\Phi_{\mathrm{Red}}(t,t_{\mathrm i})
& = &
e^{\Omega_{\mathrm{Red}}(t,t_{\mathrm i})} \\
& = & e^{\lambda \Omega_{1}(t,t_{\mathrm i}) + \lambda^2 \Omega_{2}(t,t_{\mathrm i}) + \lambda^3 \Omega_{3}(t,t_{\mathrm i}) + \; \ldots} \qquad 
\end{eqnarray}
the Magnus expansion for $\Omega_{\mathrm{Red}}(t,t_{\mathrm i})$ then gives (see Appendix \ref{App:Magnus})
\begin{equation}
\Omega_{1} = \lambda
\int_{t_{\mathrm i}}^t
\exd s\, 
\dot{\Phi}_1(s,t_{\mathrm{i}})  = \Phi_{1}(t,t_\mathrm{i}) \\
\end{equation}
as well as
\begin{eqnarray}
\Omega_{2}
& = &
\int_{t_{\mathrm i}}^t
\exd s\,
\dot{\Phi}_2^{\mathrm c}(s,t_{\mathrm i})
+ 
\int_{t_{\mathrm i}}^t
\exd s
\int_{t_{\mathrm i}}^s
\exd u\, \frac{\left[
\dot{\Phi}_1(s,t_{\mathrm i}),
\dot{\Phi}_1(u,t_{\mathrm i})
\right]}{2} \notag \\
& = &
\Phi_2^{\mathrm c}(t,t_{\mathrm i})
 - \frac{1}{2} \int_{t_{\mathrm i}}^t
\exd s
\int_{t_{\mathrm i}}^s
\exd u\, \big[ [H_{\mathrm{MF}}(s),H_{\mathrm{MF}}(u)] , \, \cdot \, \big] \notag \\
& = &
\Phi_2^{\mathrm c}(t,t_{\mathrm i})
-i
\big[
H_{\mathrm{MF}}^{(2)}(t,t_{\mathrm i}),\,\cdot\,
\big]
\notag
\\
& = &
\mathcal K_2(t,t_{\mathrm i}) 
\label{Omega2_MF}
\end{eqnarray}
where we used Eq.~(\ref{K2split_MF}).
Thus the Redfield exponent through second order is exactly as expected
\begin{eqnarray}
\Phi_{\mathrm{Red}}(t,t_{\mathrm i}) & = & e^{\lambda \Phi_{1}(t,t_{\mathrm i}) + \lambda^2 \mathcal{K}_{2}(t,t_{\mathrm i}) + \lambda^3 \Omega_{3}(t,t_{\mathrm i}) + \; \ldots} \qquad 
\end{eqnarray}
which agrees with the exponent of the CPTP Magnus map
(\ref{MagnusMap_MF}) to second-order. The first difference between the two maps therefore appears generically at order $\lambda^3$. At this next order, the Magnus expansion gives
\begin{small}
\begin{eqnarray}
\Omega_3
& = &
\int_{t_{\mathrm i}}^t
\exd s
\int_{t_{\mathrm i}}^s
\exd u\,
\frac{ 
\left[
\dot{\Phi}_1(s,t),
\dot{\Phi}_2^{\mathrm c}(u)
\right]
+
\left[
\dot{\Phi}_2^{\mathrm c}(s),
\dot{\Phi}_1(u)
\right]}{2} \label{Omega3_MF}
\\
&&
+
\frac{1}{6}
\int_{t_{\mathrm i}}^t
\exd s
\int_{t_{\mathrm i}}^s
\exd u
\int_{t_{\mathrm i}}^u
\exd v\,
\bigg\{
\left[
\dot{\Phi}_1(s),
\left[
\dot{\Phi}_1(u),
\dot{\Phi}_1(v)
\right]
\right]
\notag
\\
&&\hspace{35mm}
+
\left[
\dot{\Phi}_1(v),
\left[
\dot{\Phi}_1(u),
\dot{\Phi}_1(s)
\right]
\right]
\bigg\} \; . \notag
\end{eqnarray}
\end{small}\ignorespaces
This suggests the natural generalization of the diagnostic
(\ref{epMag_def}),
\begin{equation}
\epsilon_{\mathrm{Mag}}^{\mathrm{MF}}(t,t_{\mathrm i})
:=
\frac{
\lambda^3
\left|
\Omega_3(t,t_{\mathrm i})
\right|
}{
\left|
\lambda\Phi_1(t,t_{\mathrm i})
+
\lambda^2\mathcal K_2(t,t_{\mathrm i})
\right|
} \; .
\label{epMag_MF}
\end{equation}
Again $\epsilon_{\mathrm{Mag}}^{\mathrm{MF}}\ll1$ means that the leading correction to the Redfield exponent beyond that retained by the CPTP Magnus map is parametrically small. As in the main text, this should be regarded as a practical diagnostic rather than a rigorous convergence criterion.

When the environment one-point functions vanish, $\dot{\Phi}_1=0$, and so $\Omega_3=0$. The leading correction to the Redfield exponent then occurs at order $\lambda^4$, and the criterion reduces to Eq.~(\ref{epMag_def}) used throughout the main text.

\end{appendix}

\clearpage 

\bibliography{referencesRed.bib}

\end{document}